\documentclass[]{interact}

\usepackage{epstopdf}
\usepackage[caption=false]{subfig}
\usepackage[shortlabels]{enumitem}

\usepackage{natbib}
\bibpunct[, ]{(}{)}{;}{a}{}{,}
\renewcommand\bibfont{\fontsize{10}{12}\selectfont}

\theoremstyle{plain}

\theoremstyle{definition}

\theoremstyle{remark}

\begin{document}

\articletype{}

\title{Uncovering dynamics between SARS-CoV-2 wastewater concentrations and community infections via Bayesian spatial functional concurrent regression}

\author{
\name{Thomas Y. Sun\textsuperscript{a}\thanks{CONTACT Thomas Y. Sun. Email: tys1@alumni.rice.edu}, Julia C. Schedler\textsuperscript{a,b}, Daniel R. Kowal\textsuperscript{a,c}, Rebecca Schneider\textsuperscript{d}, Lauren B. Stadler\textsuperscript{e}, Loren Hopkins\textsuperscript{a,d} and Katherine B. Ensor\textsuperscript{a}}
\affil{\textsuperscript{a}Department of Statistics, Rice University, Houston, Texas, USA; \textsuperscript{b}Department of Statistics, California Polytechnic State University, San
Luis Obispo, San
Luis Obispo, California, USA; \textsuperscript{c}Department of Statistics and Data Science, Cornell University, Ithaca, New York, USA; \textsuperscript{d}Houston Health Department, Houston, Texas, USA; \textsuperscript{e}Department of Civil and Environment Engineering, Rice University, Houston, Texas, USA.}
}

\maketitle

\begin{abstract}
Monitoring wastewater concentrations of SARS-CoV-2 yields a low-cost, noninvasive method for tracking disease prevalence and provides early warning signs of upcoming outbreaks in the serviced communities. There is tremendous clinical and public health interest in understanding the exact dynamics between wastewater viral loads and infection rates in the population. As both data sources may contain substantial noise and missingness, in addition to spatial and temporal dependencies, properly modeling this relationship must address these numerous complexities simultaneously while providing interpretable and clear insights. We propose a novel Bayesian functional concurrent regression model that accounts for both spatial and temporal correlations while estimating the dynamic effects between wastewater concentrations and positivity rates over time. We explicitly model the time lag between the two series and provide full posterior inference on the possible delay between spikes in wastewater concentrations and subsequent outbreaks. We estimate a time lag likely between 5 to 11 days between spikes in wastewater levels and reported clinical positivity rates. Additionally, we find a dynamic relationship between wastewater concentration levels and the strength of its association with positivity rates that fluctuates between outbreaks and non-outbreaks.
\end{abstract}

\begin{keywords}
COVID-19 data; factor models; sparse functional data
\end{keywords}

\section{Introduction}
\label{sec:intro-ww}

Wastewater-based epidemiology (WBE) has emerged as a cost-efficient approach for surveilling disease prevalence within populations, notably finding extensive application in monitoring viral pathogens such as SARS-CoV-2. As infected individuals shed the virus through fecal matter, taking wastewater samples from sewage and monitoring the levels of viral matter within the samples allows for detecting the prevalence of infections in the nearby community. Numerous studies have found a significant correlation between SARS-CoV-2 wastewater data and clinical testing data for COVID-19 \citep{peccia_measurement_2020, acosta_longitudinal_2022, kaya_correlation_2022, hopkins_citywide_2023}.

There are several advantages to pairing wastewater-based surveillance with clinical data. First, there is a high resource and cost burden of large-scale testing efforts. Clinical data may be insufficient and exhibit selection bias \citep{accorsi_how_2021, symanski_population-based_2021} as these rely on voluntary efforts from individuals to seek testing. Instead, WBE offers a less biased, convenient approach to understanding population health by monitoring wastewater. Furthermore, wastewater viral loads have been shown to be a leading indicator of positive clinical cases and ICU admissions \citep{daoust_covid-19_2021,  cavany_inferring_2022, wu_sars-cov-2_2022, hopkins_citywide_2023}, offering early detection of COVID-19 outbreaks in the populations served by the wastewater catchments. This ``lead time'' metric has been a valued as a warning to potential viral outbreaks in the surrounding communities. In locales where clinical testing data is sparse or unreliable, wastewater surveillance provides a cost-effective alternative to track the population infection dynamics and may be used to predict upcoming outbreaks and their severity. Monitoring wastewater focused on specific regions or facilities can be used to track disease prevalence in those localized populations \citep{morvan_analysis_2022, wolken_wastewater_2023, ensor_online_2024}. Thus, there is strong motivation to characterize and quantify this relationship between wastewater-based viral loads and community testing metrics.  

In this paper, we consider the problem of modeling the association and lead time between wastewater viral concentration measurements and the clinical positivity rate across multiple sites in a major metropolitan area. The noisy, irregular measurements, missingness patterns, and spatial and temporal dependencies within the two sources of data present a unique modeling challenge. However, previous statistical analyses on similar wastewater data have been limited in their modeling assumptions and flexibility, and are inadequate to handle the multiple modeling challenges simultaneously. We propose a novel Bayesian functional concurrent regression approach that can account for the sparse, noisy measurements and dependencies. To illustrate these challenges and exemplify the benefits of our proposed approach, we first introduce the motivating dataset.

\subsection{Houston clinical testing data and wastewater measurements}
\label{sec:data-ww}

We introduce our motivating dataset of interest to illustrate the various sophistications at hand. Measurements of SARS-CoV-2 RNA viral loads in the wastewater (WW) were taken every week at wastewater treatment plants (WWTPs) across Houston, Texas. These WWTPs each cover their own service area in Houston and collectively serve over 2.1 million people. Meanwhile, daily COVID-19 testing and positivity rate (PR) data, defined as the ratio between the number of positive cases and number of tests, is recorded within each WWTP catchment area. Days with less than five positive cases in a given WWTP area are considered as missing. Both WW and PR data are available from March 2020 to October 2023. 

\begin{figure}
\centering
    \subfloat[]{\label{fig:wwtpex1}\includegraphics[width=.55\columnwidth]{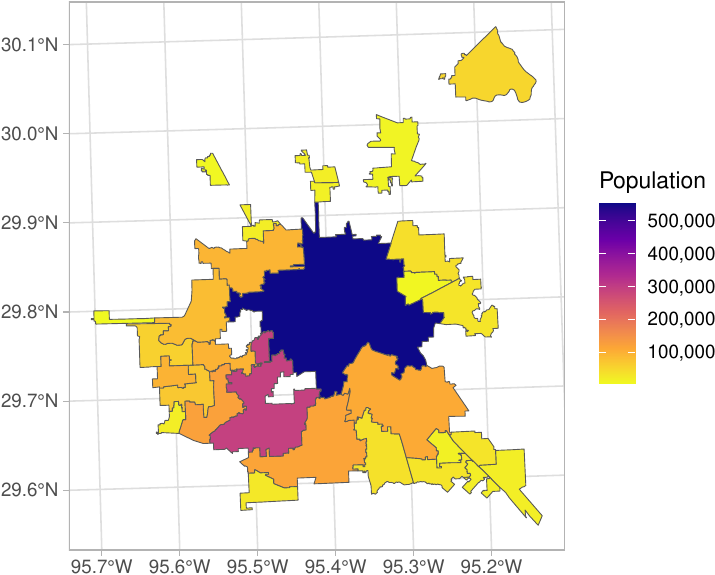}} \\
    \medskip
    \subfloat[]{\label{fig:wwtpex2}\includegraphics[width=.80\columnwidth]{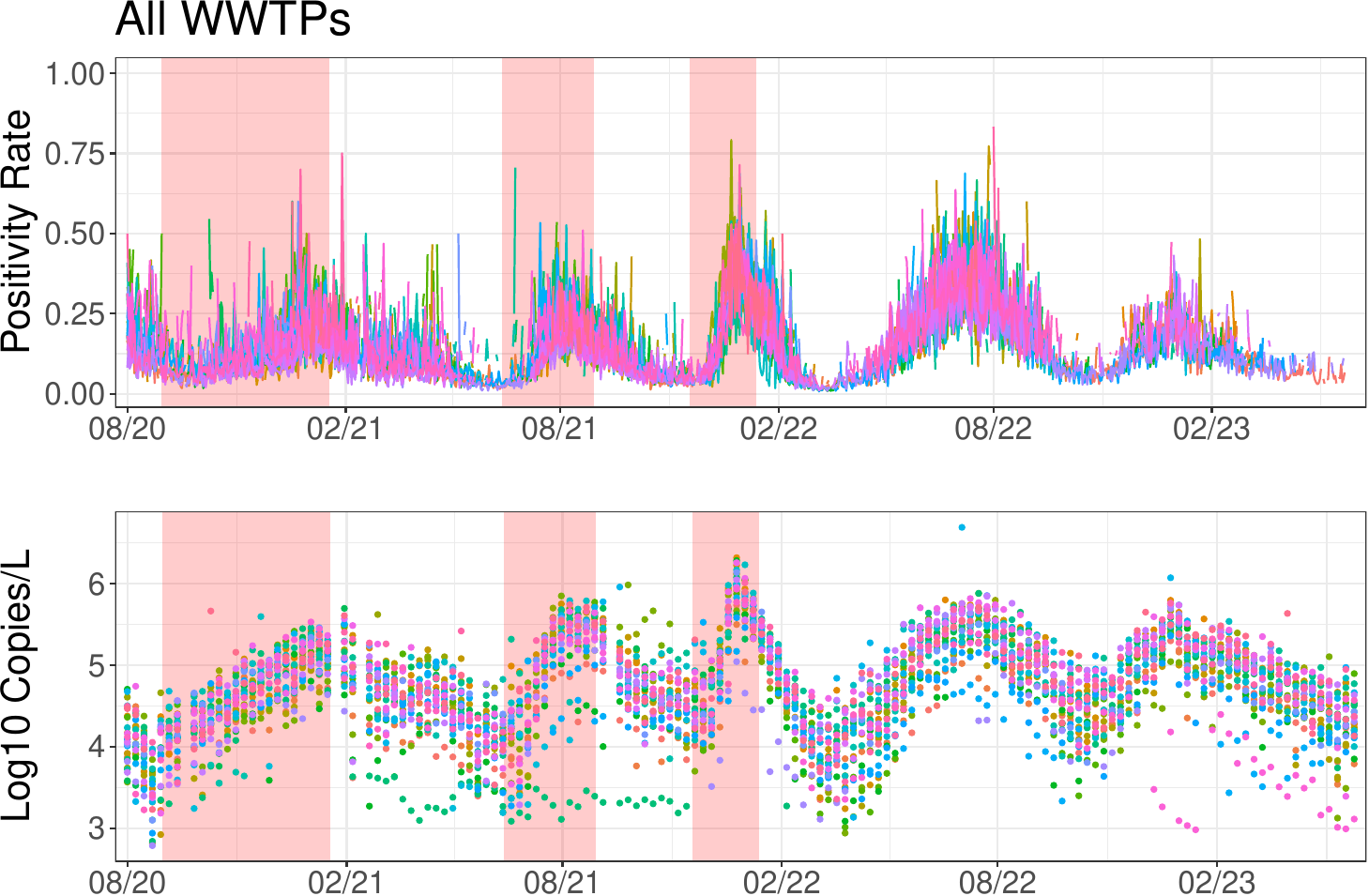}} \\
    \caption{(a) Service areas of 28 largest WWTPs in Houston and their population. (b) Positivity rate data (top) and wastewater SARS-CoV-2 viral concentration measurements in log copies per liter (bottom) of the 28 WWTPs over time. Alpha, Delta, and Omicron variant waves marked in red.}
    \label{fig:wwtpex}
\end{figure}

Figure 1 shows the geographical coverage areas and populations of the 28 largest WWTPs in Houston, along with the 28 time series of the wastewater measurements and positivity rates for each of those WWTPs. Both measurements are clearly correlated across time. There is also substantial missingness in both datasets. The wastewater measurements are taken on a weekly basis, and thus the viral concentration levels on days between sampling dates are unknown. On the other hand, information on positivity rates from clinical testing results became increasingly unavailable over time, eventually halting completely. Additionally, WWTPs servicing smaller populations may have periods of missingness during non-outbreak periods when positive results were low. From a glance, it is apparent that spikes in the positivity rates correspond to increases in the wastewater concentrations around the same time period. This indicates some relationship between the two series, but the exact relationship is not immediately clear, nor is the relationship necessarily constant across time. Also, the wastewater values likely lead the PR values by several days. Previous studies have corroborated such a relationship through correlation and cross-correlation analyses \citep{li_correlation_2023}. Furthermore, the similarities in the shapes and timing of outbreaks across each location suggests the presence of spatial correlation.

\subsection{Statistical methods}
\label{sec:prevmethods}

Several statistical challenges are apparent when modeling wastewater and inferring connection with clinical data. Clear time dependence is exhibited in both data, but numerous analyses assume wastewater measurements over time to be independent \citep{ daoust_covid-19_2021, acosta_longitudinal_2022, holm_sars-cov-2_2022} and compare differences between sewershed measurements using a simple group-level difference of means or pairwise correlations. Failure to account for time dependence may underestimate variability and inadequately capture the temporal dynamics of the wastewater measurements, where clear trends and patterns in the data exist. Similarly, when multiple measurements are taken across many locations, incorporating spatial dependence provides correct standard errors and inference. It also allows for information from nearby locations to be shared and thus helps informs us of the disease dynamics in areas, like smaller WWTPs, where the testing data is sparse. 

Temporal dynamics have previously been handled via autoregressive or regression-based models \citep{peccia_sars-cov-2_2020, stadler_wastewater_2020, cao_forecasting_2021, fitzgerald_site_2021,  vallejo_modeling_2022, jeng_application_2023}, those in the broad class of SIR or SEIR models \citep{fazli_wastewater-based_2021, mcmahan_covid-19_2021}, or machine learning approaches \citep{li_data-driven_2021, lai_time_2023}. \citet{li_spatio-temporal_2023} provide a spatiotemporal model for wastewater dynamics, but do not link this to actual clinical data. Furthermore, as \citet{torabi_wastewater-based_2023} notes in a review of spatiotemporal WBE models, even fewer approaches account for the measurement error. Although these approaches utilize well-established models, they may have considerable challenges in simultaneously incorporating measurement error, missing data, and additional sources of variability within the inference, while still providing clear, interpretable results.

Furthermore, there is interest in quantifying the lead time between wastewater concentrations of SARS-CoV-2 and clinical COVID-19 results \citep{peccia_sars-cov-2_2020, stadler_wastewater_2020,  kaplan_aligning_2021,  galani_sars-cov-2_2022}, but results can be highly variable with different statistical models reporting estimates anywhere between 0-14 days \citep{olesen_making_2021}. These estimates have generally been derived from exploratory cross correlation or correlation analyses, which requires either using the noisy raw estimates or a pre-smoothed version of the WW and PR values, without an actual inferential model on estimating the lead value. 

We provide a comprehensive statistical methodology to study the wastewater dynamics in relation to the positivity rate while addressing the previously highlighted data and statistical issues in a unified framework. Functional data analysis (FDA) offers a distinctive path forward. FDA naturally characterizes a sample of observations of an underlying process across a continuous domain. Here we can represent both the wastewater values and daily positivity rate data over time at each WWTP as functions or curves. As each unit of observation is considered a single, smoothed curve and can be represented in a low rank fashion, typically via basis functions, taking an FDA approach conveniently accounts for autocorrelation of measurements between time points. Thus we may model the temporal relationships between the two functional datasets, the wastewater curves and positivity rate curves, in a dynamic manner that reflects the evolving conditions of the pandemic over time. 

We advance a Bayesian functional concurrent regression model for noisy, sparsely observed functional data with spatial correlations and time lag. The functional concurrent regression model can provide dynamic prediction of trajectories of the response curve based on the current value of the predictor curve. In this case, the wastewater measurements are a known leading indicator for the clinical positivity rate, and hence we incorporate an unknown time lag parameter to be estimated under the Bayesian model. The resulting model can provide dynamic predictions of positivity rates based on the historic values of the wastewater concentrations. Both functional data are modeled via functional factor models, which provide data-driven, smooth estimates from the sparse, noisy data sources. Under our unified, fully Bayesian framework, we also account for the uncertainty about the weekly wastewater measurements in our final inference through a  measurement error model. To our knowledge, FDA methods have not yet been applied in the COVID-19 WBE literature. In previous analyses of Houston wastewater epidemiology, regression spline methods have been utilized in \citet{stadler_wastewater_2020}, \citet{hopkins_citywide_2023}, and \citet{ensor_online_2024} which are convenient for their functional simplicity and responsiveness, and we aim to step further in this direction with a full FDA approach. \citet{dai_statistical_2022} and later \cite{dai_bayesian_2024} predict daily cases from wastewater measurements in a Bayesian functional Poisson regression. However, they crucially do not incorporate any spatial dependence and assume the lag parameter and wastewater measurements are pre-estimated or fixed. 

Our model finds that the association between wastewater concentrations and positivity rates is highly dynamic over time, where the effect size varies depending on the specific outbreak. We also find that the lead time of wastewater is likely between 5-11 days ahead of the positivity rate. The novel method uncovers some unique insights for wastewater-based surveillance that may be informative for public health measures.

The article is written as follows. We illustrate our proposed model, priors, and MCMC algorithm in Section \ref{sec:model-ww}. Section \ref{sec:app-ww} contains the results from the application. We conclude with a discussion in Section \ref{sec:disc-ww}.

\section{Proposed Model}
\label{sec:model-ww}

There are several key characteristics of the particular dataset to be addressed simultaneously. Broadly, the primary challenges are:

\begin{enumerate}[(a)]
    \item modeling the relationship between the two functional datasets \textit{dynamically},
    \item accounting for time lag between WW measurements and PR,
    \item incorporating potential spatial dependence between WWTPs,
    \item estimating underlying functions in the presence of missing values, noise, and/or measurement error while propagating any uncertainty from such estimation.
\end{enumerate}
We first introduce the proposed model before highlighting how each specific modeling decision and component explicitly addresses each of these characteristics. Let ${y}_i(\tau)$, $i = 1, \dots, n$ be a sample of functions representing the positivity rate at time $\tau$ at WWTP (location) $i$, where  $\tau \in \mathcal{T}$. Paired with this, let $x_{i}(\tau)$, $i = 1, \dots, n$ be the measurement of wastewater viral concentrations at time $\tau$ and location $i$. Consider the following functional concurrent regression model:
\begin{equation}
\begin{gathered}
y_{i}(\tau) = X_i(\tau- \Delta){\gamma}(\tau ) + {\theta}_{i}(\tau) + \epsilon_{y_i}\left(\tau\right),  \quad  \epsilon_{yi}(\tau) \overset{iid}{\sim} N(0, \sigma^2_{\epsilon_y})\label{eq:fcr0a}
\end{gathered}
\end{equation} 
\begin{equation}
\begin{gathered}
x_{i}(\tau) = {X}_i(\tau) + \epsilon_{x_i}\left(\tau\right) ,  \quad  \epsilon_{xi}(\tau) \overset{iid}{\sim} N(0, \sigma^2_{\epsilon_x}) \label{eq:fcr0b}
\end{gathered}
\end{equation} 
\begin{equation}
\begin{gathered}
{X}_i(\tau) = \mu(\tau) + \alpha_i(\tau) \label{eq:fcr0c}
\end{gathered}
\end{equation} 
\noindent
For (a), the association between the functional response $y_i(\cdot)$ and the functional predictor $X_i(\cdot)$ in equation \eqref{eq:fcr0a} is through the ``concurrent" effect $\gamma(\cdot)$. In a standard functional concurrent regression where $\Delta = 0$, this implies that the effect of the predictor on the response $y_i(\tau)$ at $\tau$ only depends on the value of $X_i$ at exactly time $\tau$. However, viral shedding measured through wastewater occurs prior to noticeable symptoms of COVID-19 infection and the wastewater measurements are a known leading indicator of the positivity rates. Thus for (b), the time lag parameter $\Delta$ is included to incorporate this information. Now, the PR $y_i(\tau)$ at time $\tau$ depends on the value of $X_i(\tau - \Delta)$ at some past time $\tau - \Delta$. Functional concurrent regression allows for estimating a \textit{dynamic} effect $\gamma(\cdot)$ where the exact relationship between WW concentrations and PR can vary over time. This can be considered as a subset of functional regression models, where $y_i(\cdot)$ can depend $X_i(\cdot)$ over an arbitrary range of time points across the domain $\mathcal{T}$. However, we consider the concurrent model in this application for the stability and interpretability of the estimated effect. Furthermore, we also include functional random effects $\{ \theta_i(\tau) \}$ which account for smooth within curve variations of $\{y_i(\tau) \}$ and, for (c), will be used to incorporate the spatial correlations across $i$ (see Section \ref{sec:priors-spatial}), and $\epsilon_{yi}(\tau)$ are the independent, noisy errors.

For (d), in addition to the concurrent regression \eqref{eq:fcr0a}, we also include the measurement error model \eqref{eq:fcr0b}. That is, we do not assume to actually observe the true wastewater concentrations ${X}_i(\tau)$, but rather a noisy realization of them, ${x}_i(\tau)$, with added independent errors $\epsilon_{xi}$. This designation accounts for the additional uncertainty from the wastewater measurements, which is often ignored in the literature \citep{torabi_wastewater-based_2023}. 

Also for (d), the underlying smooth measurements ${X}_i(\tau)$ are modeled via a population level mean function $\mu(\tau)$ and random effect function $\alpha_i(\tau)$. Moreover, \eqref{eq:fcr0b} can be considered a functional mixed effects model where information is pooled across curves. This information borrowing provides stability to each curve estimate as compared to modeling each curve individually. Such models are popularly used for sparsely observed functional data (see \citealp{shi_analysis_1996, james_principal_2000, yao_functional_2005}, among others). Further, although the WW data are observed on a weekly basis versus the daily PR data, our functional approach elegantly resolves this irregularity in sampling frequencies, as we may estimate the underlying continuous functions between measured time points and incorporate the associated uncertainty within our inference.

We model the functional coefficient $\gamma(\tau)$, as well the random effects $\theta_i(\tau)$, $\mu(\tau)$, and $\alpha_i(\tau)$ via the following basis function expansions:

\begin{equation}
\begin{gathered}
\gamma(\tau) = \sum^{P}_{p=1} b_{p}(\tau)\gamma_p\label{eq:basis-gamma}
\end{gathered}
\end{equation} 
\begin{equation}
\begin{gathered}
\theta_i(\tau) = \sum^{L}_{\ell=1} g_{\ell}(\tau) \theta_{\ell, i}\label{eq:basis-theta}
\end{gathered}
\end{equation} 
\begin{equation}
\begin{gathered}
X_i(\tau) = \sum^{K}_{k=1}f_k(\tau_{})\beta_{k,i} = \sum^{K}_{k=1}f_k(\tau_{})(\mu_k + \alpha_{k,i})\label{eq:basis-x}
\end{gathered}
\end{equation} 
\noindent
where $\{b_p(\cdot)\}$, $\{g_{\ell} (\cdot)\}$, and  $\{f_k (\cdot)\}$ are the respective basis functions with coefficients $\{\gamma_p\}$, $\{\theta_{\ell, i}\}$, and $\{\beta_{k, i}\}$.

Further considerations about the basis function representations are necessary in the case of noisy functional data with high missingness such as the WW and PR curves. Estimating underlying smooth curves as in \eqref{eq:basis-theta}--\eqref{eq:basis-x} may be difficult when there are insufficient observed values. The positivity rates are not observed late in 2023 as testing efforts declined and eventually stopped. Assuming a known and given basis may produce highly unstable estimates or may be infeasible altogether unless the number of basis functions is set to be small. But, this would result in a highly inflexible model where the densely observed portions of the functions are underfit. Popular methods for sparse functional data include functional principal components analysis (FPCA) approaches \citep{yao_functional_2005, peng_geometric_2009, cederbaum_functional_2016}. Indeed, these have been used for previous functional concurrent regression models \citep{senturk_functional_2010,leroux_dynamic_2018} as they provide a parsimonious eigenfunction basis which can be estimated from the data. However, using pre-estimated eigenfunctions does not account for the uncertainty within the regression model, and such methods have been found to underestimate total variability \citep{goldsmith_corrected_2013}.

Based on this background, we use choose to model the basis functions $\{f_k (\cdot)\}$ and $\{g_{\ell} (\cdot)\}$ as unknown using functional factor models in a Bayesian approach. Functional factor models provide a flexible, low-rank representation of the noisy, sparse measurements and adaptively selects the basis functions according to available data. Bayesian functional factor models and the closely related Bayesian FPCA approaches \citep{suarez_bayesian_2017} have been studied under a wide context of regression models \citep{montagna_bayesian_2012, goldsmith_generalized_2015, kowal_semiparametric_2023} including functional concurrent regression \citep{goldsmith_variable_2017}. The functions are modeled as $f_k(\tau) = \sum^{H}_{h = 1}w_h(\tau)\psi_{h,k}$ and $g_{\ell}(\tau) = \sum^{J}_{j=1} v_j(\tau)\phi_{j,\ell}$ respectively where $\{w_h \}$ and $\{v_j\}$ are sets of known basis functions with unknown coefficients $\{\psi_{h,k}\}$ and $\{\phi_{j,\ell}\}$. For the known basis functions $\{ w_h(\tau) \}$, we use low rank thin plate splines (LR-TPS), where we follow the construction based on \cite{kowal_dynamic_2021}, and diagonalize the associated penalty matrix then orthogonalize the basis functions, a beneficial strategy for improved computational and MCMC efficiency \citep{sun_ultra-efficient_2024} and ensuring positive definiteness of the prior precision matrix. The basis functions $\{ v_j(\tau) \}$ are constructed using Demmler-Reinsch basis functions, which orthogonalize a penalized B-spline basis with knots at every time point \citep{wand_semiparametric_2008}. Our factor model ensures that the level of smoothness is adaptively selected while properly accounting for the underlying uncertainty about the unknown loading curves $\{f_k (\cdot)\}$ and $\{g_{\ell} (\cdot)\}$. We assume $\{b_p(\cdot)\}$ are known basis functions via B-splines with a penalty on the second-order differences.

\subsection{Prior and model specifications}
\label{sec:priors}

Since we only observe the functions at a discrete set of time points, we rewrite \eqref{eq:fcr0a}--\eqref{eq:fcr0b} in discrete form. Let $\boldsymbol{\tau} = \{\tau_1, \dots,\tau_{t}, \dots,  \tau_M \} \in \mathcal{T}$ be a common grid of time points where each $y_i$ and $x_i$ is observed on a subset of this grid. Since $y_i$ and $x_i$ may be observed on different time points, let $\boldsymbol{x}_i = \boldsymbol{x}_i(\boldsymbol{\tau}^x_{i})$ be a vector representing the wastewater measurements along a discrete set of time points $\boldsymbol{\tau}^x_{i}$ of size $m^x_i$, and likewise for the testing data $\boldsymbol{y}_i = \boldsymbol{y}_i(\boldsymbol{\tau}^y_{i})$ and size $m^y_i$, where $\boldsymbol{\tau}^x_{i}  \subseteq \boldsymbol{\tau}$ and $\boldsymbol{\tau}^y_{i}  \subseteq \boldsymbol{\tau} \ \forall i$.  
\begin{equation}
\begin{gathered}
\bm y_i(\boldsymbol{\tau}^y_{i}) =  \text{diag}(\bm \gamma(\boldsymbol{\tau}^y_{i}))\bm X_i(\boldsymbol{\tau}^y_{i} - \Delta) + \bm \theta_i(\boldsymbol{\tau}^y_{i}) + \bm \epsilon_{yi} ,   \quad  \boldsymbol{\epsilon}_{yi} \overset{indep}{\sim} N_{m^y_i}(\boldsymbol{0}, \sigma^2_{\epsilon_y}\boldsymbol{I}_{m^y_i}),,\label{eq:fcrd}
\end{gathered}
\end{equation}  
\begin{equation}
\begin{gathered}
\boldsymbol{x}_i(\boldsymbol{\tau}^x_{i}) = \bm X_i(\boldsymbol{\tau}^x_{i}) + \boldsymbol{\epsilon}_{xi} =  \boldsymbol{F}_i(\boldsymbol{\tau}^x_{i})\boldsymbol{\beta}_{i} + \boldsymbol{\epsilon}_{xi},  \quad  \boldsymbol{\epsilon}_{xi} \overset{indep}{\sim} N_{m^x_i}(\boldsymbol{0}, \sigma^2_{\epsilon_x}\boldsymbol{I}_{m^x_i}),
\label{eq:memd}
\end{gathered}
\end{equation}
\noindent
where $\bm F = (\bm f_1^{\prime}, \dots, \bm f_K^{\prime})^{\prime}$ where $\bm f_{k} = (f_{k}(\tau_1), \dots, f_{k}(\tau_M ))^{\prime}$, and similarly $\bm G = (\bm g_1^{\prime}, \dots, \bm g_L^{\prime})^{\prime}$, $\bm B  = (\bm b_1^{\prime}, \dots, \bm b_P^{\prime})^{\prime}$, $\bm W = (\bm w_1^{\prime}, \dots, \bm w_H^{\prime})^{\prime}$, and $\bm V  = (\bm v_1^{\prime}, \dots, \bm v_J^{\prime})^{\prime}$, such that $\bm \gamma_i = \bm B_i \bm \gamma$, $\boldsymbol{F}_i = \boldsymbol{W}_i\boldsymbol{\Psi}$ and $\boldsymbol{G}_i = \boldsymbol{V}_i\boldsymbol{\Phi}$.

In the measurement error model \eqref{eq:memd} we assume that the coefficients $\{\mu_k \}$ and $\{\alpha_{k,i} \}$ follow a multiplicative gamma process (MGP) \citep{bhattacharya_sparse_2011}, which promotes identifiability of the loading curves and reduces sensitivity on the number of factors $K$. The prior for the factors is given by $\mu_k \sim N(0, \sigma^2_{\mu_k})$ and $\alpha_{k,i} \sim N(0, \sigma^2_{\alpha_{k}}/\zeta_{\alpha_{k,i}})$. Further details on the MGP hyperpriors are available in the supplement. Identifiability of the loading curves is not explicitly required in the Bayesian approach, although the loading curves are constrained to enforce orthogonality, $\bm F^{\prime}\bm F = \bm I_{K}$. For the observation error we assume $[\sigma^{-2}_{\epsilon_x}] \sim \text{Gamma}(a_{\epsilon_x}, b_{\epsilon_x})$. For the loading curves $\{ f_k \}$ we place a roughness penalty on coefficents $\{ \psi_{\ell, k} \}$  through the prior $[ \bm \psi_k | \lambda_{f_k} ] \stackrel{indep}{\sim} N(\boldsymbol{0}, \lambda_{f_k}^{-1}\bm \Omega^{-}_{\psi})$ with $\bm \Omega_{\psi}$ is a known penalty matrix on the second differences and prior precision parameter $\lambda_{f_k}^{-1/2} \overset{iid}\sim \text{Uniform}(0, 10^4)$, and similarly for the loading curves $\{g_{\ell} \}$ we let $[ \bm \phi_{\ell} | \lambda_{g_{\ell}} ] \stackrel{indep}{\sim} N(\boldsymbol{0}, \lambda_{g_{\ell}}^{-1}\bm \Omega^{-}_{\phi})$ and $\lambda_{g_{\ell}}^{-1/2} \overset{iid}\sim \text{Uniform}(0, 10^4)$.

\subsubsection{Prior for lag parameter}\label{priors-lag}
For the lag parameter, we assume that $\Delta \in \{0, \dots, \Delta_{max}\}$ and assume it follows a categorical distribution with equal probability for each possible lag. We set $\Delta_{max}$ to 21 days, which implies that the lag between WW and PR measurements is anywhere between 0 and up to 3 weeks. The resulting posterior distribution for $\Delta$ will also follow a categorical distribution and provide probability values for each possible lag based on model fit. For the concurrent regression coefficient we assume that $\bm \gamma = (\gamma_1, \dots, \gamma_P)^{\prime} \sim N_P(\bm 0 , \lambda_{\gamma}^{-1} \bm \Omega_{\gamma}^{-})$ where $\lambda_{\gamma}^{-1/2}$ follows an uninformative flat prior and $\bm \Omega_{\gamma}$ is a known second order differences penalty matrix to promote smoothness. 

\subsubsection{Spatial dependence prior}\label{sec:priors-spatial}
To incorporate spatial dependence in the random effect functions let $\bm \theta_{\ell} = (\theta_{{\ell},1}, \dots, \theta_{{\ell}, n})^{\prime}$. We assume that $\bm \theta_{\ell}$ is spatially correlated with $\bm \theta_{\ell} \sim N_n(\bm 0, \sigma^{2}_{\theta_{\ell}} \bm Q^{-1} )$ where $\bm Q = (\text{diag}(\bm D\bm1) - \bm D)$. $\bm D$ is a spatial weight matrix and $\bm 1$ is an n-dimensional vector of 1s, while $\sigma^{-2}_{\theta_{\ell}} \sim \text{Gamma}(.01, .01)$ controls the scale. This implies that each $\bm \theta_{\ell} $ follows a conditionally autoregressive model where $\bm D$ controls the influence of nearby neighbors on each coefficient.

Although many choices of $\bm D$ are available given areal spatial data, the particular irregularities in the WWTP coverage areas as shown in Figure \ref{fig:wwtpex1} indicates the need for a robust distance metric to define the spatial relationships. Namely, we use the extended Hausdorff distance \citep{min_extended_2007}, which is specifically designed for defining distances on geometric spatial objects. The Hausdorff distance is defined on two sets and measures the maximum separation between the sets by finding the supremum of all infimum distances between points of one set to the other set, while the extended Hausdorff distance replaces the supremum with any desired quantile of distances. In our application we use the median Hausdorff distance, which offers a robust construction of similarity between two WWTP coverage areas \citep{schedler_spatiotemporal_2021}. We then define $\bm D$ as a spatial weight matrix containing the 10 nearest neighbors of each WWTP as defined by extended Hausdorff distance, with symmetry enforced to define a proper CAR model.

\subsection{MCMC algorithm}
\label{sec:mcmc-ww}

We obtain posterior inference for all model parameters in Section \ref{sec:priors} via a Gibbs sampling algorithm. The main components of the sampler consist of cycling through (1) the regression coefficient $\bm \gamma$, (2) the spatial random effects coefficients $\{ \bm \theta_p \}$, (3) the loading curve parameters $\{ \phi_{j,\ell}\}$, (4) the functional factor components $\{\mu_k \}$, $\{\alpha_{k,i} \}$ and (5) their associated loading curve parameters $\{ \psi_{h,k}\}$, (6) lag parameter $\Delta$, and (7) the variance components. With the exception of (2), we condition only on values at the observed time points $\boldsymbol{\tau}^x_{i}$ and $\boldsymbol{\tau}^y_{i}$, which importantly reduces autocorrelation and increases stability in the algorithm. Otherwise, imputed draws can be easily be taken from the posterior predictive distribution. Details about the full conditional distributions are available in the supplementary material. 

\section{Dynamic association between SARS-CoV-2 wastewater levels and COVID-19 reported positivity rates in Houston}
\label{sec:app-ww}

We now apply our functional concurrent regression model to the motivating dataset in Section \ref{sec:data-ww}. Our data consists of wastewater and positivity rate measurements between August 3rd, 2020 to June 3rd, 2023 from the 28 largest WWTP coverage areas based on population size. We run 100,000 MCMC iterations with the initial 40,000 discarded as burn-in. Our MCMC diagnostics showed favorable convergence of chains with high effective sample size.

\begin{figure}[!ht]
\centering
\includegraphics[width=.8\textwidth]{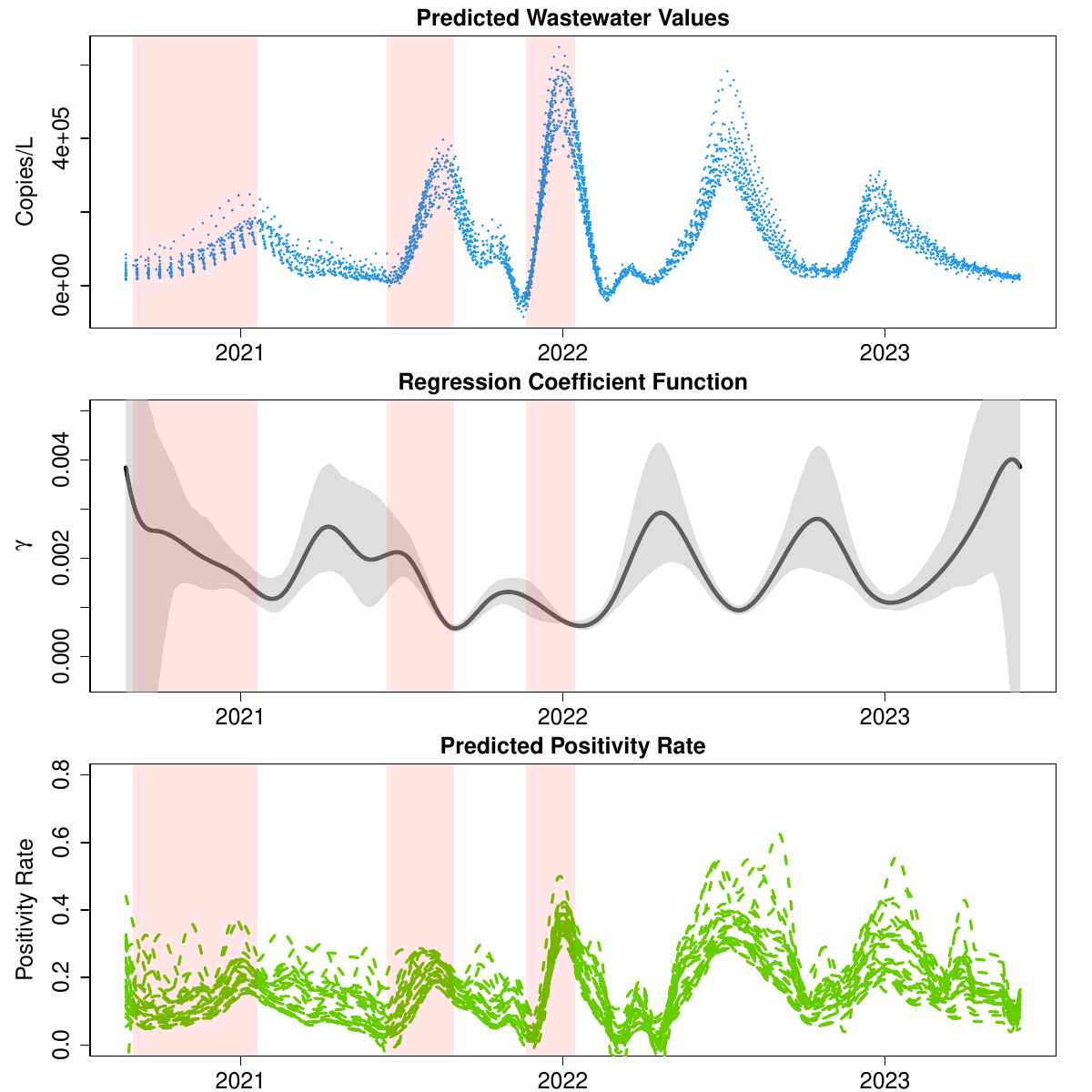}
\caption{Predicted wastewater curves from 28 WWTPs, the estimate of regression coefficient function $\gamma(\tau)$ with 95\% credible intervals (middle), and the predicted positivity rate curves from all 28 WWTPs (bottom). The Alpha, Delta, and Omicron variant waves are marked in the red bars. The relationship between wastewater measurements and positivity rates are highly dynamic, with the effect $\gamma(\tau)$ sharply decreasing during outbreak periods.}
  \label{fig:fitplots}
\end{figure}

\begin{figure}[t]
\centering
\includegraphics[width=.67\textwidth]{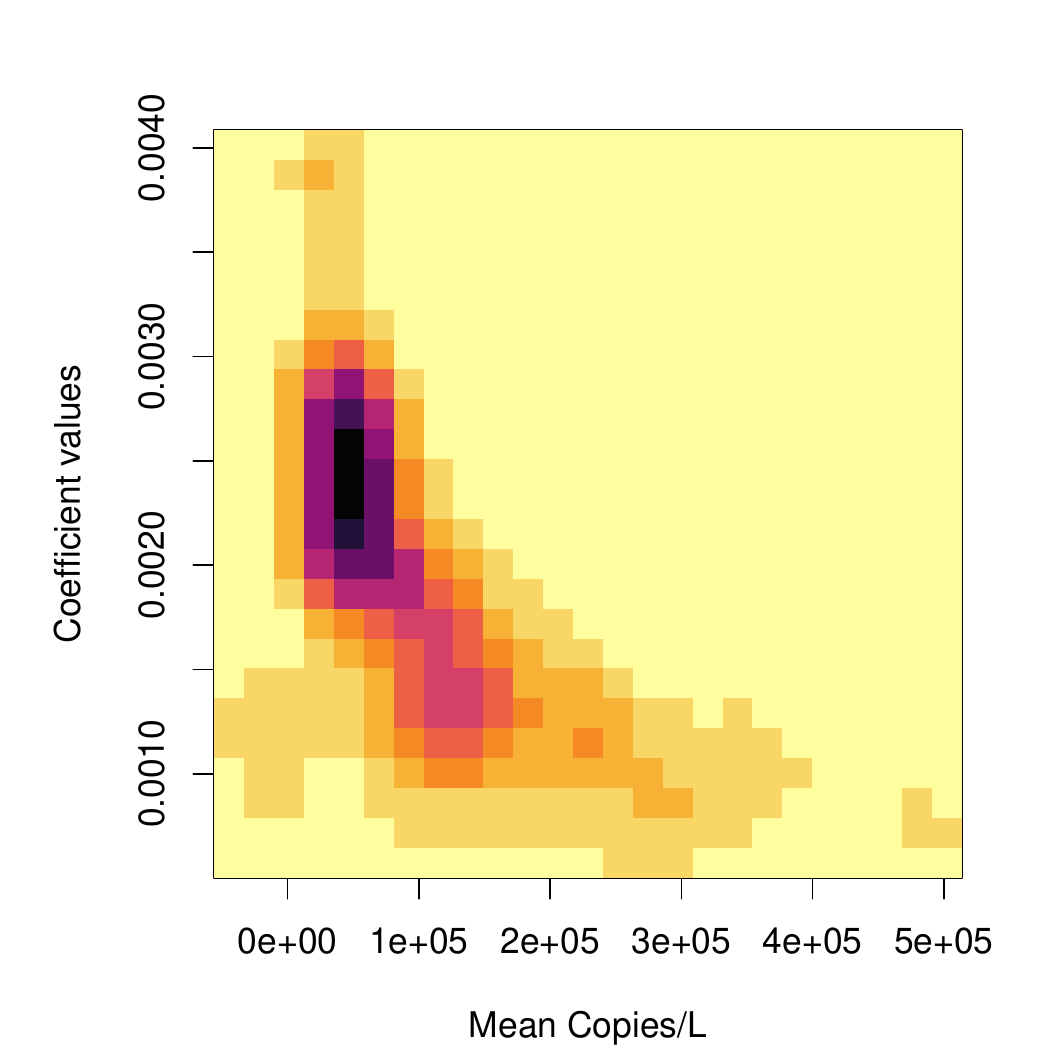}
\caption{Heatmap of the bivariate density between the estimated mean population wastewater levels $\mu(\tau)$ and estimated regression coefficient $\gamma(\tau)$ ignoring the time index. The density suggests an inverse relationship between the level of mean wastewater concentration and the regression coefficient values.}
  \label{fig:heatmap2}
\end{figure}

Figure \ref{fig:fitplots} shows the fitted wastewater curves, fitted positivity rate curves, and the posterior mean and 95\% pointwise credible interval of the regression coefficient function $\gamma(\tau)$. The red bars mark the periods of the Alpha, Delta, and Omicron SARS-CoV-2 variant waves \citep{stadler_wastewater_2020}. The regression coefficient effect has marked changes across time, indicating that this  bivariate relationship is highly dynamic. In particular, we observe an inverse association between wastewater levels and the value of the regression coefficient. During non-outbreak periods, any increase in the level of wastewater concentrations when overall levels are low will have a relatively strong association with the positivity rate. The opposite is true during periods of high concentrations of wastewater viral load, where the estimated coefficient is lower value. Based on the credible bands, periods of peak wastewater levels also have the lowest uncertainty in the regression coefficient, whereas variability widened during flatter periods. During the beginning period up to the Alpha wave shows higher uncertainty levels. This may partly be an issue with data quality as testing efforts were still nascent. This is also likely the case for the latter period of 2023.

To corroborate the inverse relationship between the wastewater concentration levels and the regression coefficient function $\gamma(\tau)$, we calculate the posterior mean of the population level mean function $\mu(\tau)$ and compare these values with the values of $\gamma(\tau)$ by looking at the bivariate density ignoring the time index. Figure \ref{fig:heatmap2} shows a heatmap of these bivariate values. From this, we see a clear inverse trend where the coefficient values amass around 0.002 and 0.003 at low wastewater concentration levels whereas these values fall closer to 0.001 at high wastewater concentration levels. Thus, while monitoring the wastewater levels can provide clear insights to the relative severity of disease outbreaks in the served population, this relationship should be differentiated between outbreak and non-outbreak periods. The association with positivity rate is lower in terms of expected value but has much higher precision during waves, and thus any relative increase in concentration at a WWTP compared to other WWTPs in this time period has meaningful impact on the relative severity of the outbreak in that served community.

\begin{figure}[t]
\centering
\includegraphics[width=.87\textwidth]{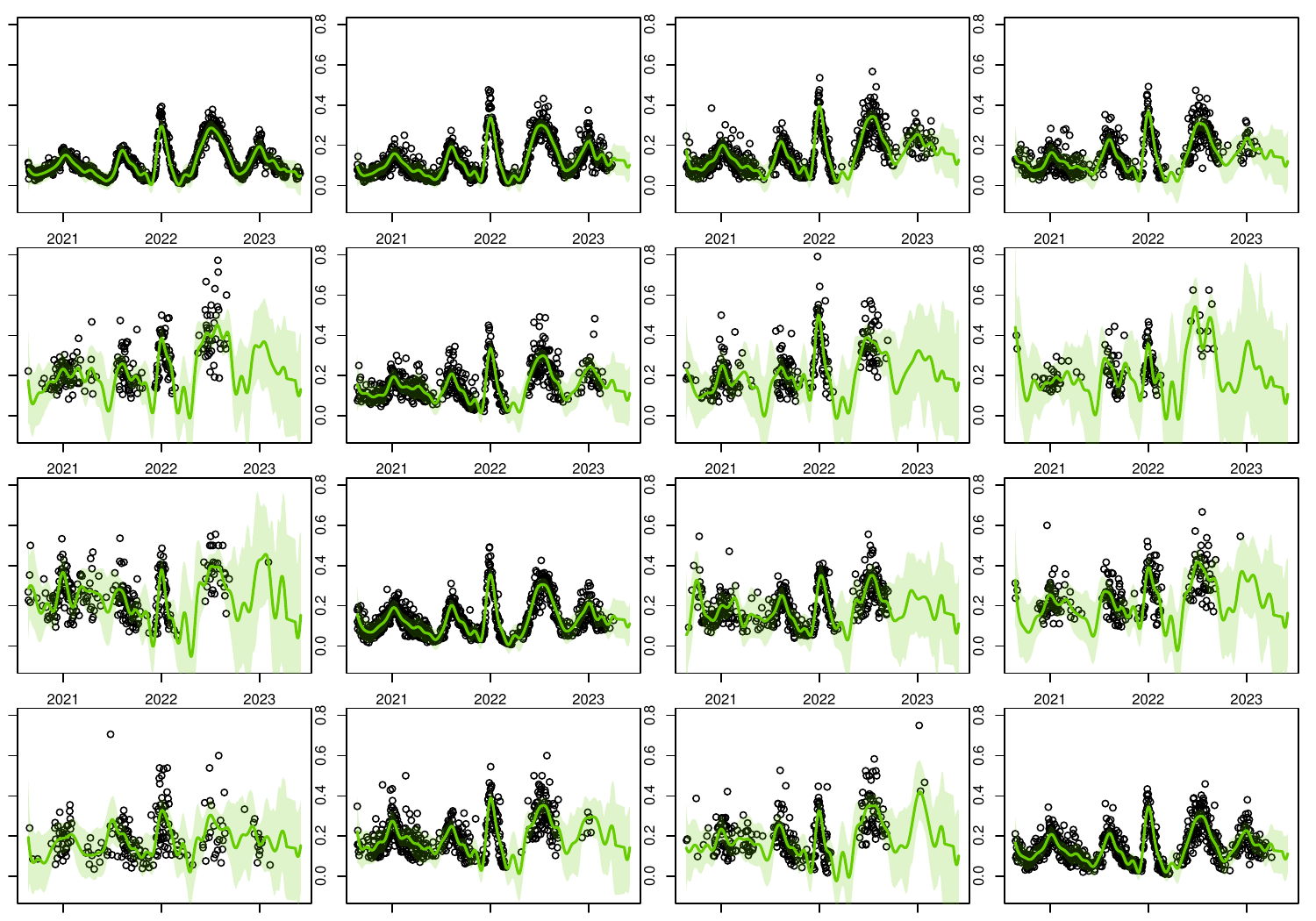}
\caption{Raw daily reported COVID-19 positivity rate values (dots) with fitted posterior mean curves (solid lines) and 95\% pointwise credible intervals (shaded regions) of the largest 16 WWTPs by population. The functional regression model can provide predictions and uncertainty to missing portions of PR curves at locations with insufficient data.}
  \label{fig:estprpost}
\end{figure}

\begin{figure}[t]
\centering
\includegraphics[width=.87\textwidth]{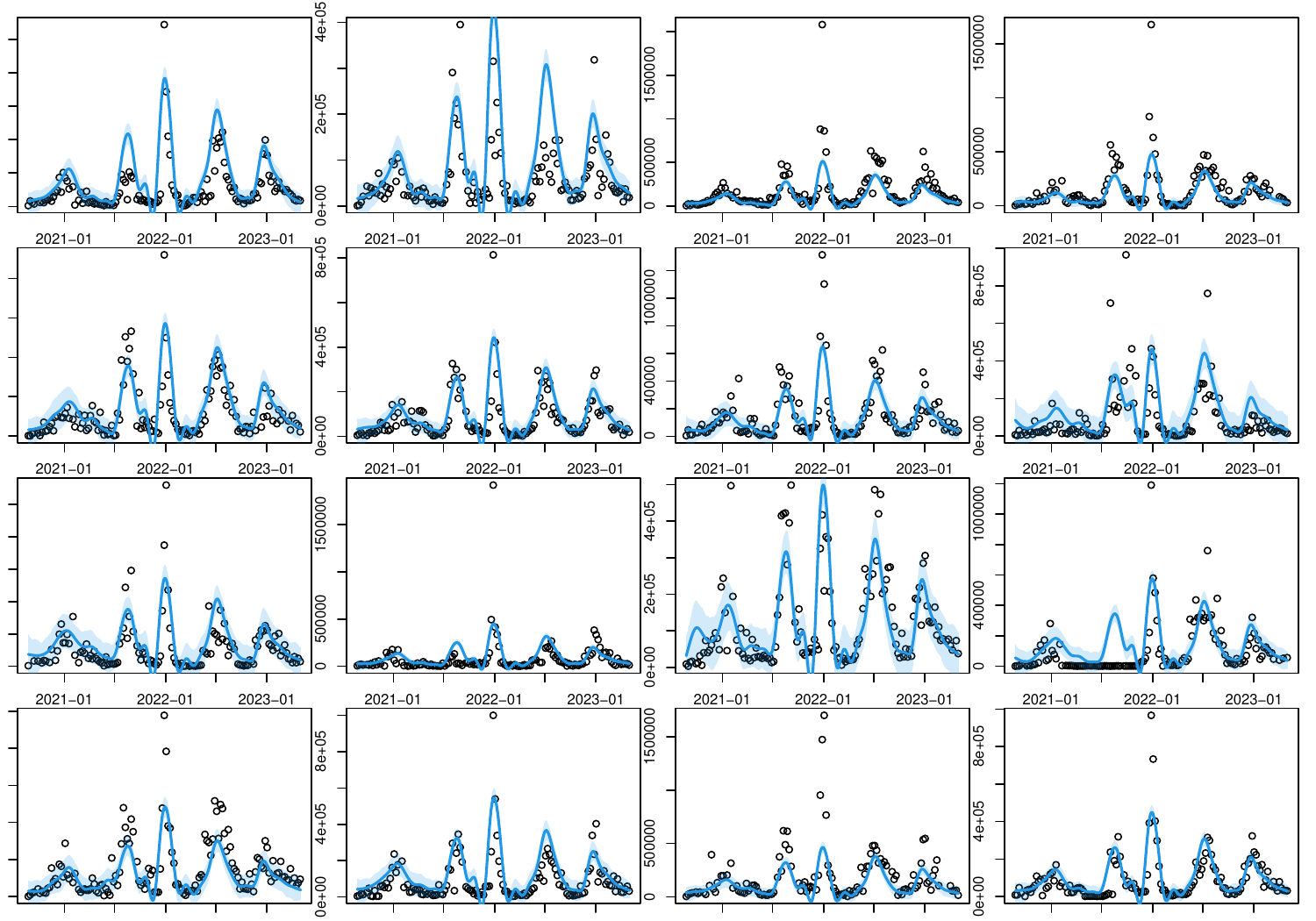}
\caption{Raw weekly SARS-CoV-2 wastewater measurements (dots) with fitted posterior mean curves (solid lines) and 95\% pointwise credible intervals (dashed lines) of the largest 16 WWTPs by population. The Bayesian measurement error model provides estimates and uncertainty at times in between weekly measurements and incorporates information jointly from other model parameters.}
  \label{fig:estwwpost}
\end{figure}

Figure \ref{fig:estprpost} shows the raw positivity rate values of the largest 16 WWTPs along with the fitted posterior mean curve and 95\% pointwise credible intervals. Likewise, Figure \ref{fig:estwwpost} shows the raw wastewater measurements with corresponding fitted mean curve and 95\% pointwise credible intervals. The remaining WWTPs are shown in the supplementary material. The raw PR values clearly show that testing efforts were greatest during the early periods of the pandemic, where most WWTPs have at least some measurements, especially during the initial waves. But in the last year or so of available data, many curves are completely unobserved as public testing waned. Importantly, our model is able to provide estimates to the sparsely observed regions through the wastewater measurements by borrowing information from nearby service areas via the spatial CAR component. Thus, the method captures the trends and waves in COVID-19 prevalence even in regions and time periods of low testing data, along with the associated uncertainty.

\begin{figure}[h]
\centering
\includegraphics[width=.8\textwidth]{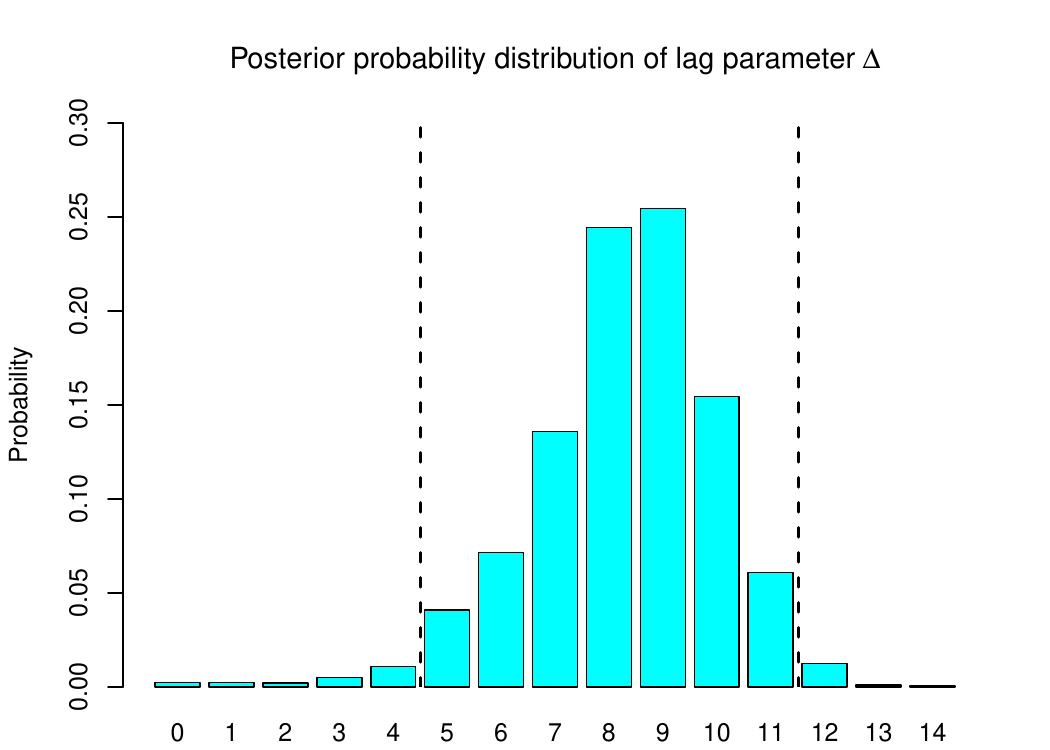}
\caption{Posterior probability distribution of the lag parameter $\Delta$ and 95\% HPD interval (vertical dashed lines), denoting the estimated lead time in days between the wastewater measurement and the clinical positivity rate.}
  \label{fig:lagpost}
\end{figure}

Figure \ref{fig:lagpost} shows the posterior distribution of the lag parameter $\Delta$. The 95\% highest posterior density (HPD) credible interval, as shown by the dashed lines, indicates that the most likely lag values fall between 5 to 11 days, while the most probable lag is between 8 or 9 days. This implies that spikes in the wastewater SARS-CoV-2 concentrations likely lead the positivity rate data by several days, most likely between a little over a week. Despite including up to 21 days of possible values, the model did not find that a lead time beyond 14 days was probable. These results generally follow the results from previous statistical models and correlation analyses. However, our inference importantly provides an entire distribution of possible lag values. Therefore under conservative estimates, the spike in wastewater levels may indicate an incoming COVID-19 wave as soon as 5 days in advance.

\section{Discussion}
\label{sec:disc-ww}

In this work we demonstrated the advantages of taking an FDA approach to modeling of wastewater-based epidemiological data. The noise, irregularity and sparsity of both the wastewater and positivity rate measurements are readily handled by FDA models when carefully constructed. We show that with FDA, the relationship between the two measurements can be flexibly modeled via functional regression, uncovering new insights on their dynamics. Furthermore, our fully Bayesian approach directly incorporates uncertainty from unknown sources to key model parameters, including the time lag parameter and the measurement error and missingness from the wastewater measurements. Incorporating spatial dependence not only accounts for the additional autocorrelation present, but also allows for borrowing of information across WWTPs. If testing data is sparse in a lightly populated region, the spatial functional model smooths and incorporates data from nearby regions. We assume the spatial dependence is constant in time. This work highlights the importance of not only wastewater-based surveillance but also utilizing multiple wastewater sampling sites to provide a more informative and granular understanding of disease prevalence.

While we primarily investigated the clinical positivity rates as our outcome of interest, studying other metrics of COVID-19 prevalence, such as ICU hospitalization rates, may be compelling to further research. Using reported postivity rate data may be criticized as unreliable and inconsistent as mass testing efforts declined. Incorporating information about the number of tests taken each day may account for this source of measurement error. As forecasting future levels of COVID-19 prevalence is of crucial scientific interest, future works for developing forecasting models with functional concurrent regression may be worth exploring.

\section*{Disclosure Statement}
No potential conflict of interest was reported by the author(s).

\section*{Author Attribution and Funding}
Sun developed and applied the functional Bayesian methods and provided the original drat. Schedler developed and implemented the spatial structure and assisted with analysis and editing. Kowal oversaw the development of the methodology and implementation. Schneider manages the analytics for Houston Health Department WBE program. Stadler leads the wastewater epidemiology lab at Rice, providing expertise and SARS-CoV-2 RNA measurements. Hopkins leads the wastewater epidemiology program for Houston Health Department. Ensor proposed and oversaw the technical devleopment of this work, and leads the analytics effort for the Houston WBE program. This work was supported by the Centers for Disease Control and Prevention (ELC-ED grant no. 6NU50CK000557-01-05 and
ELC-CORE grant no. NU50CK000557).

\newpage

\bibliographystyle{tfcse}
\bibliography{mybibww032624}

\begin{thebibliography}{50}
\providecommand{\natexlab}[1]{#1}
\providecommand{\url}[1]{\normalfont{#1}}
\providecommand{\urlprefix}{Available from: }

\bibitem[Accorsi et~al.(2021)]{accorsi_how_2021}
Accorsi~EK, Qiu~X, Rumpler~E, Kennedy-Shaffer~L, Kahn~R, Joshi~K, Goldstein~E, Stensrud~MJ, Niehus~R, Cevik~M, et~al. 2021. How to detect and reduce potential sources of biases in studies of {SARS}-{CoV}-2 and {COVID}-19. European Journal of Epidemiology. 36(2):179--196.  [accessed 2024-10-24]. \urlprefix\url{http://link.springer.com/10.1007/s10654-021-00727-7}.

\bibitem[Acosta et~al.(2022)]{acosta_longitudinal_2022}
Acosta~N, Bautista~MA, Waddell~BJ, McCalder~J, Beaudet~AB, Man~L, Pradhan~P, Sedaghat~N, Papparis~C, Bacanu~A, et~al. 2022. Longitudinal {SARS}-{CoV}-2 {RNA} wastewater monitoring across a range of scales correlates with total and regional {COVID}-19 burden in a well-defined urban population. Water Research. 220:118611.  [accessed 2023-08-18]. \urlprefix\url{https://doi.org/10.1016/j.watres.2022.118611}.

\bibitem[Bhattacharya and Dunson(2011)]{bhattacharya_sparse_2011}
Bhattacharya~A, Dunson~DB. 2011. Sparse {Bayesian} infinite factor models. Biometrika. 98(2):291--306.  [accessed 2021-06-11]. \urlprefix\url{https://academic.oup.com/biomet/article-lookup/doi/10.1093/biomet/asr013}.

\bibitem[Cao and Francis(2021)]{cao_forecasting_2021}
Cao~Y, Francis~R. 2021. On forecasting the community-level {COVID}-19 cases from the concentration of {SARS}-{CoV}-2 in wastewater. Science of The Total Environment. 786:147451.  \urlprefix\url{https://www.sciencedirect.com/science/article/pii/S0048969721025225}.

\bibitem[Cavany et~al.(2022)]{cavany_inferring_2022}
Cavany~S, Bivins~A, Wu~Z, North~D, Bibby~K, Perkins~TA. 2022. Inferring {SARS}-{CoV}-2 {RNA} shedding into wastewater relative to the time of infection. Epidemiology and Infection. 150:e21.  [accessed 2024-03-18]. \urlprefix\url{https://www.cambridge.org/core/product/identifier/S0950268821002752/type/journal\_article}.

\bibitem[Cederbaum et~al.(2016)]{cederbaum_functional_2016}
Cederbaum~J, Pouplier~M, Hoole~P, Greven~S. 2016. Functional linear mixed models for irregularly or sparsely sampled data. Statistical Modelling. 16(1):67--88.  [accessed 2023-05-27]. \urlprefix\url{http://journals.sagepub.com/doi/10.1177/1471082X15617594}.

\bibitem[Dai et~al.(2024)]{dai_bayesian_2024}
Dai~X, Acosta~N, Lu~X, Hubert~CRJ, Lee~J, Frankowski~K, Bautista~MA, Waddell~BJ, Du~K, McCalder~J, et~al. 2024. A {Bayesian} framework for modeling {COVID}‐19 case numbers through longitudinal monitoring of {SARS}‐{CoV}‐2 {RNA} in wastewater. Statistics in Medicine. 43(6):1153--1169.  [accessed 2024-03-17]. \urlprefix\url{https://onlinelibrary.wiley.com/doi/10.1002/sim.10009}.

\bibitem[Dai et~al.(2022)]{dai_statistical_2022}
Dai~X, Champredon~D, Fazil~A, Mangat~CS, Peterson~SW, Mejia~EM, Lu~X, Chekouo~T. 2022. Statistical framework to support the epidemiological interpretation of {SARS}-{CoV}-2 concentration in municipal wastewater. Scientific Reports. 12(1):13490.  [accessed 2024-03-17]. \urlprefix\url{https://www.nature.com/articles/s41598-022-17543-y}.

\bibitem[D'Aoust et~al.(2021)]{daoust_covid-19_2021}
D'Aoust~PM, Towhid~ST, Mercier~E, Hegazy~N, Tian~X, Bhatnagar~K, Zhang~Z, Naughton~CC, MacKenzie~AE, Graber~TE, et~al. 2021. {COVID}-19 wastewater surveillance in rural communities: {Comparison} of lagoon and pumping station samples. Science of The Total Environment. 801:149618.  [accessed 2023-08-18]. \urlprefix\url{https://doi.org/10.1016/j.scitotenv.2021.149618}.

\bibitem[Ensor et~al.(2024)]{ensor_online_2024}
Ensor~KB, Schedler~JC, Sun~T, Schneider~R, Mulenga~A, Wu~J, Stadler~LB, Hopkins~L. 2024. Online trend estimation and detection of trend deviations in sub-sewershed time series of {SARS}-{CoV}-2 {RNA} measured in wastewater. Scientific Reports. 14(1):5575.  [accessed 2024-03-18]. \urlprefix\url{https://www.nature.com/articles/s41598-024-56175-2}.

\bibitem[Fazli et~al.(2021)]{fazli_wastewater-based_2021}
Fazli~M, Sklar~S, Porter~MD, French~BA, Shakeri~H. 2021. Wastewater-{Based} {Epidemiological} {Modeling} for {Continuous} {Surveillance} of {COVID}-19 {Outbreak}. In: 2021 {IEEE} {International} {Conference} on {Big} {Data} ({Big} {Data}); Dec; Orlando, FL, USA. IEEE. p. 4342--4349.  [accessed 2023-09-11]. \urlprefix\url{https://ieeexplore.ieee.org/document/9671543/}.

\bibitem[Fitzgerald et~al.(2021)]{fitzgerald_site_2021}
Fitzgerald~SF, Rossi~G, Low~AS, McAteer~SP, O’Keefe~B, Findlay~D, Cameron~GJ, Pollard~P, Singleton~PTR, Ponton~G, et~al. 2021. Site {Specific} {Relationships} between {COVID}-19 {Cases} and {SARS}-{CoV}-2 {Viral} {Load} in {Wastewater} {Treatment} {Plant} {Influent}. Environmental Science \& Technology. 55(22):15276--15286. Publisher: American Chemical Society;  \urlprefix\url{https://doi.org/10.1021/acs.est.1c05029}.

\bibitem[Galani et~al.(2022)]{galani_sars-cov-2_2022}
Galani~A, Aalizadeh~R, Kostakis~M, Markou~A, Alygizakis~N, Lytras~T, Adamopoulos~PG, Peccia~J, Thompson~DC, Kontou~A, et~al. 2022. {SARS}-{CoV}-2 wastewater surveillance data can predict hospitalizations and {ICU} admissions. Science of The Total Environment. 804:150151.  [accessed 2024-03-24]. \urlprefix\url{https://linkinghub.elsevier.com/retrieve/pii/S0048969721052281}.

\bibitem[Goldsmith et~al.(2013)]{goldsmith_corrected_2013}
Goldsmith~J, Greven~S, Crainiceanu~C. 2013. Corrected {Confidence} {Bands} for {Functional} {Data} {Using} {Principal} {Components}. Biometrics. 69(1):41--51.  [accessed 2024-02-13]. \urlprefix\url{https://academic.oup.com/biometrics/article/69/1/41-51/7490897}.

\bibitem[Goldsmith and Schwartz(2017)]{goldsmith_variable_2017}
Goldsmith~J, Schwartz~JE. 2017. Variable selection in the functional linear concurrent model. Statistics in Medicine. 36(14):2237--2250.  [accessed 2024-03-05]. \urlprefix\url{https://onlinelibrary.wiley.com/doi/10.1002/sim.7254}.

\bibitem[Goldsmith et~al.(2015)]{goldsmith_generalized_2015}
Goldsmith~J, Zipunnikov~V, Schrack~J. 2015. Generalized {Multilevel} {Function}-on-{Scalar} {Regression} and {Principal} {Component} {Analysis}. Biometrics. 71(2):344--353.  [accessed 2024-02-19]. \urlprefix\url{https://academic.oup.com/biometrics/article/71/2/344-353/7511367}.

\bibitem[Holm et~al.(2022)]{holm_sars-cov-2_2022}
Holm~RH, Mukherjee~A, Rai~JP, Yeager~RA, Talley~D, Rai~SN, Bhatnagar~A, Smith~T. 2022. {SARS}-{CoV}-2 {RNA} abundance in wastewater as a function of distinct urban sewershed size. Environmental Science: Water Research \& Technology. 8(4):807--819. Publisher: The Royal Society of Chemistry;  [accessed 2023-08-18]. \urlprefix\url{https://doi.org/10.1039/D1EW00672J}.

\bibitem[Hopkins et~al.(2023)]{hopkins_citywide_2023}
Hopkins~L, Persse~D, Caton~K, Ensor~K, Schneider~R, McCall~C, Stadler~LB. 2023. Citywide wastewater {SARS}-{CoV}-2 levels strongly correlated with multiple disease surveillance indicators and outcomes over three {COVID}-19 waves. Science of The Total Environment. 855:158967.  [accessed 2023-08-11]. \urlprefix\url{https://doi.org/10.1016/j.scitotenv.2022.158967}.

\bibitem[James et~al.(2000)]{james_principal_2000}
James~GM, Hastie~TJ, Sugar~CA. 2000. Principal component models for sparse functional data. Biometrika. 87(3):587--602.  [accessed 2023-11-03]. \urlprefix\url{https://academic.oup.com/biomet/article-lookup/doi/10.1093/biomet/87.3.587}.

\bibitem[Jeng et~al.(2023)]{jeng_application_2023}
Jeng~HA, Singh~R, Diawara~N, Curtis~K, Gonzalez~R, Welch~N, Jackson~C, Jurgens~D, Adikari~S. 2023. Application of wastewater-based surveillance and copula time-series model for {COVID}-19 forecasts. Science of The Total Environment. 885:163655.  [accessed 2023-09-11]. \urlprefix\url{https://linkinghub.elsevier.com/retrieve/pii/S0048969723022751}.

\bibitem[Kaplan et~al.(2021)]{kaplan_aligning_2021}
Kaplan~EH, Wang~D, Wang~M, Malik~AA, Zulli~A, Peccia~J. 2021. Aligning {SARS}-{CoV}-2 indicators via an epidemic model: application to hospital admissions and {RNA} detection in sewage sludge. Health Care Management Science. 24(2):320--329.  \urlprefix\url{https://doi.org/10.1007/s10729-020-09525-1}.

\bibitem[Kaya et~al.(2022)]{kaya_correlation_2022}
Kaya~D, Falender~R, Radniecki~T, Geniza~M, Cieslak~P, Kelly~C, Lininger~N, Sutton~M. 2022. Correlation between {Clinical} and {Wastewater} {SARS}-{CoV}-2 {Genomic} {Surveillance}, {Oregon}, {USA}. Emerging Infectious Diseases. 28(9):1906--1908.  [accessed 2023-08-18]. \urlprefix\url{https://doi.org/10.3201/eid2809.220938}.

\bibitem[Kowal(2021)]{kowal_dynamic_2021}
Kowal~DR. 2021. Dynamic {Regression} {Models} for {Time}-{Ordered} {Functional} {Data}. Bayesian Analysis. 16(2).  [accessed 2023-11-03]. \urlprefix\url{https://projecteuclid.org/journals/bayesian-analysis/volume-16/issue-2/Dynamic-Regression-Models-for-Time-Ordered-Functional-Data/10.1214/20-BA1213.full}.

\bibitem[Kowal and Canale(2023)]{kowal_semiparametric_2023}
Kowal~DR, Canale~A. 2023. Semiparametric {Functional} {Factor} {Models} with {Bayesian} {Rank} {Selection}. Bayesian Analysis. 18(4).  [accessed 2024-02-26]. \urlprefix\url{https://projecteuclid.org/journals/bayesian-analysis/volume-18/issue-4/Semiparametric-Functional-Factor-Models-with-Bayesian-Rank-Selection/10.1214/23-BA1410.full}.

\bibitem[Lai et~al.(2023)]{lai_time_2023}
Lai~M, Cao~Y, Wulff~SS, Robinson~TJ, McGuire~A, Bisha~B. 2023. A time series based machine learning strategy for wastewater-based forecasting and nowcasting of {COVID}-19 dynamics. Science of The Total Environment. 897:165105.  [accessed 2023-09-11]. \urlprefix\url{https://linkinghub.elsevier.com/retrieve/pii/S0048969723037282}.

\bibitem[Leroux et~al.(2018)]{leroux_dynamic_2018}
Leroux~A, Xiao~L, Crainiceanu~C, Checkley~W. 2018. Dynamic prediction in functional concurrent regression with an application to child growth. Statistics in Medicine. 37(8):1376--1388.  [accessed 2023-11-28]. \urlprefix\url{https://onlinelibrary.wiley.com/doi/10.1002/sim.7582}.

\bibitem[Li et~al.(2023{\natexlab{a}})]{li_spatio-temporal_2023}
Li~G, Denise~H, Diggle~P, Grimsley~J, Holmes~C, James~D, Jersakova~R, Mole~C, Nicholson~G, Smith~CR, et~al. 2023{\natexlab{a}}. A spatio-temporal framework for modelling wastewater concentration during the {COVID}-19 pandemic. Environment International. 172:107765.  [accessed 2024-03-21]. \urlprefix\url{https://linkinghub.elsevier.com/retrieve/pii/S0160412023000387}.

\bibitem[Li et~al.(2021)]{li_data-driven_2021}
Li~X, Kulandaivelu~J, Zhang~S, Shi~J, Sivakumar~M, Mueller~J, Luby~S, Ahmed~W, Coin~L, Jiang~G. 2021. Data-driven estimation of {COVID}-19 community prevalence through wastewater-based epidemiology. Science of The Total Environment. 789:147947.  \urlprefix\url{https://www.sciencedirect.com/science/article/pii/S0048969721030187}.

\bibitem[Li et~al.(2023{\natexlab{b}})]{li_correlation_2023}
Li~X, Zhang~S, Sherchan~S, Orive~G, Lertxundi~U, Haramoto~E, Honda~R, Kumar~M, Arora~S, Kitajima~M, et~al. 2023{\natexlab{b}}. Correlation between {SARS}-{CoV}-2 {RNA} concentration in wastewater and {COVID}-19 cases in community: {A} systematic review and meta-analysis. Journal of Hazardous Materials. 441:129848.  [accessed 2024-07-31]. \urlprefix\url{https://linkinghub.elsevier.com/retrieve/pii/S0304389422016417}.

\bibitem[McMahan et~al.(2021)]{mcmahan_covid-19_2021}
McMahan~CS, Self~S, Rennert~L, Kalbaugh~C, Kriebel~D, Graves~D, Colby~C, Deaver~JA, Popat~SC, Karanfil~T, et~al. 2021. {COVID}-19 wastewater epidemiology: a model to estimate infected populations. The Lancet Planetary Health. 5(12):e874--e881.  \urlprefix\url{https://www.sciencedirect.com/science/article/pii/S2542519621002308}.

\bibitem[Min et~al.(2007)]{min_extended_2007}
Min~D, Zhilin~L, Xiaoyong~C. 2007. Extended {Hausdorff} distance for spatial objects in {GIS}. International Journal of Geographical Information Science. 21(4):459--475.  [accessed 2024-03-24]. \urlprefix\url{http://www.tandfonline.com/doi/abs/10.1080/13658810601073315}.

\bibitem[Montagna et~al.(2012)]{montagna_bayesian_2012}
Montagna~S, Tokdar~ST, Neelon~B, Dunson~DB. 2012. Bayesian {Latent} {Factor} {Regression} for {Functional} and {Longitudinal} {Data}. Biometrics. 68(4):1064--1073.  [accessed 2021-06-11]. \urlprefix\url{http://doi.wiley.com/10.1111/j.1541-0420.2012.01788.x}.

\bibitem[Morvan et~al.(2022)]{morvan_analysis_2022}
Morvan~M, Jacomo~AL, Souque~C, Wade~MJ, Hoffmann~T, Pouwels~K, Lilley~C, Singer~AC, Porter~J, Evens~NP, et~al. 2022. An analysis of 45 large-scale wastewater sites in {England} to estimate {SARS}-{CoV}-2 community prevalence. Nature Communications. 13(1):4313.  \urlprefix\url{https://doi.org/10.1038/s41467-022-31753-y}.

\bibitem[Olesen et~al.(2021)]{olesen_making_2021}
Olesen~SW, Imakaev~M, Duvallet~C. 2021. Making waves: {Defining} the lead time of wastewater-based epidemiology for {COVID}-19. Water Research. 202:117433.  [accessed 2024-03-18]. \urlprefix\url{https://linkinghub.elsevier.com/retrieve/pii/S004313542100631X}.

\bibitem[Peccia et~al.(2020{\natexlab{a}})]{peccia_measurement_2020}
Peccia~J, Zulli~A, Brackney~DE, Grubaugh~ND, Kaplan~EH, Casanovas-Massana~A, Ko~AI, Malik~AA, Wang~D, Wang~M, et~al. 2020{\natexlab{a}}. Measurement of {SARS}-{CoV}-2 {RNA} in wastewater tracks community infection dynamics. Nature Biotechnology. 38(10):1164--1167.  [accessed 2023-08-18]. \urlprefix\url{https://doi.org/10.1038/s41587-020-0684-z}.

\bibitem[Peccia et~al.(2020{\natexlab{b}})]{peccia_sars-cov-2_2020}
Peccia~J, Zulli~A, Brackney~DE, Grubaugh~ND, Kaplan~EH, Casanovas-Massana~A, Ko~AI, Malik~AA, Wang~D, Wang~M, et~al. 2020{\natexlab{b}}. {SARS}-{CoV}-2 {RNA} concentrations in primary municipal sewage sludge as a leading indicator of {COVID}-19 outbreak dynamics. medRxiv:2020.05.19.20105999. Publisher: Cold Spring Harbor Laboratory Press;  [accessed 2020-05-29]. \urlprefix\url{https://www.medrxiv.org/content/10.1101/2020.05.19.20105999v1}.

\bibitem[Peng and Paul(2009)]{peng_geometric_2009}
Peng~J, Paul~D. 2009. A {Geometric} {Approach} to {Maximum} {Likelihood} {Estimation} of the {Functional} {Principal} {Components} {From} {Sparse} {Longitudinal} {Data}. Journal of Computational and Graphical Statistics. 18(4):995--1015.  [accessed 2024-02-13]. \urlprefix\url{http://www.tandfonline.com/doi/abs/10.1198/jcgs.2009.08011}.

\bibitem[Schedler and Ensor(2021)]{schedler_spatiotemporal_2021}
Schedler~JC, Ensor~KB. 2021. A spatiotemporal case‐crossover model of asthma exacerbation in the {City} of {Houston}. Stat. 10(1):e357.  [accessed 2023-11-28]. \urlprefix\url{https://onlinelibrary.wiley.com/doi/10.1002/sta4.357}.

\bibitem[{\c{S}}entürk and Müller(2010)]{senturk_functional_2010}
{\c{S}}entürk~D, Müller~HG. 2010. Functional {Varying} {Coefficient} {Models} for {Longitudinal} {Data}. Journal of the American Statistical Association. 105(491):1256--1264.  [accessed 2024-03-23]. \urlprefix\url{https://www.tandfonline.com/doi/full/10.1198/jasa.2010.tm09228}.

\bibitem[Shi et~al.(1996)]{shi_analysis_1996}
Shi~M, Weiss~RE, Taylor~JMG. 1996. An {Analysis} of {Paediatric} {CD4} {Counts} for {Acquired} {Immune} {Deficiency} {Syndrome} {Using} {Flexible} {Random} {Curves}. Applied Statistics. 45(2):151.  [accessed 2024-02-13]. \urlprefix\url{https://www.jstor.org/stable/2986151?origin=crossref}.

\bibitem[Stadler et~al.(2020)]{stadler_wastewater_2020}
Stadler~L, Ensor~K, Clark~J, Kalvapalle~P, LaTurner~ZW, Mojica~L, Terwilliger~A, Zhuo~Y, Ali~P, Avadhanula~V, et~al. 2020. Wastewater {Analysis} of {SARS}-{CoV}-2 as a {Predictive} {Metric} of {Positivity} {Rate} for a {Major} {Metropolis}. Epidemiology. preprint.  [accessed 2023-08-18]. \urlprefix\url{http://medrxiv.org/lookup/doi/10.1101/2020.11.04.20226191}.

\bibitem[Suarez and Ghosal(2017)]{suarez_bayesian_2017}
Suarez~AJ, Ghosal~S. 2017. Bayesian {Estimation} of {Principal} {Components} for {Functional} {Data}. Bayesian Analysis. 12(2).  [accessed 2024-02-22]. \urlprefix\url{https://projecteuclid.org/journals/bayesian-analysis/volume-12/issue-2/Bayesian-Estimation-of-Principal-Components-for-Functional-Data/10.1214/16-BA1003.full}.

\bibitem[Sun and Kowal(2024)]{sun_ultra-efficient_2024}
Sun~TY, Kowal~DR. 2024. Ultra-{Efficient} {MCMC} for {Bayesian} {Longitudinal} {Functional} {Data} {Analysis}. Journal of Computational and Graphical Statistics:1--13.  [accessed 2024-11-23]. \urlprefix\url{https://www.tandfonline.com/doi/full/10.1080/10618600.2024.2362227}.

\bibitem[Symanski et~al.(2021)]{symanski_population-based_2021}
Symanski~E, Ensor~KB, Piedra~PA, Sheth~K, Caton~K, Williams~SL, Persse~D, Banerjee~D, Hopkins~L. 2021. Population-{Based} {Estimates} of {SARS}-{CoV}-2 {Seroprevalence} in {Houston}, {Texas} as of {September} 2020. The Journal of Infectious Diseases:jiab203.  [accessed 2024-10-24]. \urlprefix\url{https://academic.oup.com/jid/advance-article/doi/10.1093/infdis/jiab203/6258999}.

\bibitem[Torabi et~al.(2023)]{torabi_wastewater-based_2023}
Torabi~F, Li~G, Mole~C, Nicholson~G, Rowlingson~B, Smith~CR, Jersakova~R, Diggle~PJ, Blangiardo~M. 2023. Wastewater-based surveillance models for {COVID}-19: {A} focused review on spatio-temporal models. Heliyon. 9(11):e21734.  [accessed 2024-03-21]. \urlprefix\url{https://linkinghub.elsevier.com/retrieve/pii/S2405844023089429}.

\bibitem[Vallejo et~al.(2022)]{vallejo_modeling_2022}
Vallejo~JA, Trigo-Tasende~N, Rumbo-Feal~S, Conde-Pérez~K, López-Oriona~A, Barbeito~I, Vaamonde~M, Tarrío-Saavedra~J, Reif~R, Ladra~S, et~al. 2022. Modeling the number of people infected with {SARS}-{COV}-2 from wastewater viral load in {Northwest} {Spain}. Science of The Total Environment. 811:152334.  \urlprefix\url{https://www.sciencedirect.com/science/article/pii/S0048969721074118}.

\bibitem[Wand and Ormerod(2008)]{wand_semiparametric_2008}
Wand~MP, Ormerod~JT. 2008. {ON} {SEMIPARAMETRIC} {REGRESSION} {WITH} {O}'{SULLIVAN} {PENALIZED} {SPLINES}. Australian \& New Zealand Journal of Statistics. 50(2):179--198.  [accessed 2024-02-29]. \urlprefix\url{https://onlinelibrary.wiley.com/doi/10.1111/j.1467-842X.2008.00507.x}.

\bibitem[Wolken et~al.(2023)]{wolken_wastewater_2023}
Wolken~M, Sun~T, McCall~C, Schneider~R, Caton~K, Hundley~C, Hopkins~L, Ensor~K, Domakonda~K, Kalvapalle~P, et~al. 2023. Wastewater surveillance of {SARS}-{CoV}-2 and influenza in {preK}-12 schools shows school, community, and citywide infections. Water Research. 231:119648.  [accessed 2023-03-11]. \urlprefix\url{https://linkinghub.elsevier.com/retrieve/pii/S0043135423000830}.

\bibitem[Wu et~al.(2022)]{wu_sars-cov-2_2022}
Wu~F, Xiao~A, Zhang~J, Moniz~K, Endo~N, Armas~F, Bonneau~R, Brown~MA, Bushman~M, Chai~PR, et~al. 2022. {SARS}-{CoV}-2 {RNA} concentrations in wastewater foreshadow dynamics and clinical presentation of new {COVID}-19 cases. Science of The Total Environment. 805:150121.  \urlprefix\url{https://www.sciencedirect.com/science/article/pii/S0048969721051962}.

\bibitem[Yao et~al.(2005)]{yao_functional_2005}
Yao~F, Müller~HG, Wang~JL. 2005. Functional {Data} {Analysis} for {Sparse} {Longitudinal} {Data}. Journal of the American Statistical Association. 100(470):577--590.  [accessed 2023-11-03]. \urlprefix\url{http://www.tandfonline.com/doi/abs/10.1198/016214504000001745}.

\end{thebibliography}


\begin{thebibliography}{}

\bibitem[\protect\citeauthoryear{Jang, Gu, and Poole}{Jang et~al.}{2017}]{jang_categorical_2017}
Jang, E., S.~Gu, and B.~Poole (2017, August).
\newblock Categorical {Reparameterization} with {Gumbel}-{Softmax}.
\newblock arXiv:1611.01144 [cs, stat].

\bibitem[\protect\citeauthoryear{Kowal}{Kowal}{2021}]{kowal_dynamic_2021}
Kowal, D.~R. (2021, June).
\newblock Dynamic {Regression} {Models} for {Time}-{Ordered} {Functional} {Data}.
\newblock {\em Bayesian Analysis\/}~{\em 16\/}(2).

\bibitem[\protect\citeauthoryear{Neal}{Neal}{2003}]{neal_slice_2003}
Neal, R.~M. (2003, June).
\newblock Slice sampling.
\newblock {\em The Annals of Statistics\/}~{\em 31\/}(3).

\end{thebibliography}

\end{document}


\def\spacingset#1{\renewcommand{\baselinestretch}%
{#1}\small\normalsize} \spacingset{1}


\if0\blind
{
  \title{\bf Supplement to ``Uncovering dynamics between SARS-CoV-2 wastewater concentrations and community infections via Bayesian spatial functional concurrent regression"}
  \author{Thomas Y. Sun\hspace{.2cm}}
  \maketitle
} \fi

\if1\blind
{
  \bigskip
  \bigskip
  \bigskip
  \begin{center}
    {\LARGE\bf Title}
\end{center}
  \medskip
} \fi

\makeatletter
\renewcommand \thesection{S\@arabic\c@section}
\renewcommand\thetable{\@arabic\c@table}
\renewcommand \thefigure{\@arabic\c@figure}
\makeatother

\supplementarysection
\doublespacing
\section{Model and prior specification}\label{sec:model1-s}
The Bayesian functional concurrent regression model with measurement error is 
\begin{equation}
\begin{gathered}
y_{i}(\tau) = X_i(\tau- \Delta){\gamma}(\tau ) + {\theta}_{i}(\tau) + \epsilon_{y_i}\left(\tau\right),  \quad  \epsilon_{yi}(\tau) \overset{iid}{\sim} N(0, \sigma^2_{\epsilon_y})\label{fcr-s1}
\end{gathered}
\end{equation}
\begin{equation}
\begin{gathered}
x_{i}(\tau) = {X}_i(\tau) + \epsilon_{x_i}\left(\tau\right) = \mu(\tau) + \alpha_i(\tau) + \epsilon_{x_i}\left(\tau\right),  \quad  \epsilon_{xi}(\tau) \overset{iid}{\sim} N(0, \sigma^2_{\epsilon_x})\label{fcr-s2}
\end{gathered}
\end{equation}  
\begin{equation}
\begin{gathered}
\gamma(\tau) = \sum^{P}_{p=1} b_{p}(\tau)\gamma_p, \quad \bm \gamma = (\gamma_1, \dots, \gamma_P)^{\prime} \sim N_P(\bm 0 , \lambda_{\gamma}^{-1} \bm \Omega_{\gamma}^{-})\label{fcr-s3}
\end{gathered}
\end{equation}
\begin{equation}
\begin{gathered}
\mu(\tau) + \alpha_i(\tau) = \sum^{K}_{k=1}f_k(\tau_{})\beta_{k,i} = \sum^{K}_{k=1}f_k(\tau_{})(\mu_k + \alpha_{k,i}), \quad \mu_k \sim N(0, \sigma^2_{\mu_k}), \quad \alpha_{k,i} \sim N(0, \sigma^2_{\alpha_{k}}/\zeta_{\alpha_{k,i}})\label{fcr-s4}
\end{gathered}
\end{equation}
\begin{equation}
\begin{gathered}
f_k(\tau) = \sum^{H}_{h = 1}w_h(\tau)\psi_{h,k}, \quad [ \bm \psi_k | \lambda_{f_k} ] \stackrel{indep}{\sim} N(\boldsymbol{0}, \lambda_{f_k}^{-1}\bm \Omega^{-}_{\psi}) \label{fcr-s5}
\end{gathered}
\end{equation}
\begin{equation}
\begin{gathered}
\theta_i(\tau) = \sum^{L}_{\ell=1} g_{\ell}(\tau) \theta_{\ell, i}, \quad \bm \theta_{\ell} \sim N_n(\bm 0, \sigma^{2}_{\theta_{\ell}} \bm Q^{-1} )\label{fcr-s6}
\end{gathered}
\end{equation}
\begin{equation}
\begin{gathered}
g_{\ell}(\tau) = \sum^{J}_{j=1} v_j(\tau)\phi_{j,\ell}, \quad [ \bm \phi_{\ell} | \lambda_{g_{\ell}} ] \stackrel{indep}{\sim} N(\boldsymbol{0}, \lambda_{g_{\ell}}^{-1}\bm \Omega^{-}_{\phi})\label{fcr-s7}
\end{gathered}
\end{equation}
\begin{equation}
\begin{gathered}
\lambda_{f_k}^{-1/2} \overset{iid}\sim \text{Uniform}(0, 10^4), \quad \lambda_{g_{\ell}}^{-1/2} \overset{iid}\sim \text{Uniform}(0, 10^4), \quad \lambda_{\gamma}^{-1/2} \overset{iid}\sim \text{Uniform}(0, 10^4)\label{fcr-s8}
\end{gathered}
\end{equation}
\begin{equation}
\begin{gathered}
\Delta \in \{0, \dots, \Delta_{21}\} \sim \text{Categorical}\left(\frac{1}{21}\right)\label{fcr-s9}
\end{gathered}
\end{equation}
\begin{equation}
\begin{gathered}
[\sigma^{-2}_{\epsilon_x}] \sim \text{Gamma}(a_{\epsilon_x}, b_{\epsilon_x}), \quad [\sigma^{-2}_{\epsilon_y}] \sim \text{Gamma}(a_{\epsilon_y}, b_{\epsilon_y})\label{fcr-s10}
\end{gathered}
\end{equation}
\begin{equation}
\begin{gathered}
\sigma^{-2}_{\mu_k} = \prod_{k^*\leq k} \delta_{\mu_{k^*}}, \quad \delta_{\mu_{1}} \sim \text{Gamma}(a_{\mu_1}, 1), \quad \delta_{\mu_{k^*}} \sim \text{Gamma}(a_{\mu_2}, 1), \quad k^* > 1\label{fcr-s11}
\end{gathered}
\end{equation}
\begin{equation}
\begin{gathered}
\sigma^2_{\alpha_{k,i}} = \sigma^2_{\alpha_{k}}/\zeta_{\alpha_{k,i}}, \quad \zeta_{\alpha_{k,i}} \sim \text{Gamma}(\nu_{\alpha}/2, \nu_{\alpha}/2) , \quad \nu_{\alpha} \sim \text{Uniform}(2,128), \label{fcr-s12}
\end{gathered}
\end{equation}
\begin{equation}
\begin{gathered}
\sigma^{-2}_{\alpha_{k}} = \prod_{k^* \leq k} \delta_{\alpha_{k^*}}, \quad \delta_{\alpha_{1}} \sim \text{Gamma}(a_{\alpha_1}, 1), \quad \delta_{\alpha_{k^*}} \sim \text{Gamma}(a_{\alpha_2}, 1), \quad k^* > 1 \label{fcr-s13}
\end{gathered}
\end{equation}
\begin{equation}
\begin{gathered}
a_{\mu_1}, a_{\mu_2}, a_{\alpha_1}, a_{\alpha_2} \overset{iid}{\sim} \text{Gamma}(2, 1). \label{fcr-s14}
\end{gathered}
\end{equation}

\section{MCMC algorithm}\label{sec:mcmc2-s}

\begin{enumerate}
    \item Sample regression parameters:
    \begin{enumerate}
        \item     \begin{equation}
    \label{post-mu-sofr}
    \begin{gathered}
    [\bm \gamma \mid \cdots] \stackrel{}{\sim} N(Q^{-1}_{\gamma} \ell_{\gamma},  Q^{-1}_{\gamma}) \\
    Q_{\gamma} =  \bm\lambda^{-1}_{\gamma} \bm \Omega_{\gamma} +  \sigma^{-2}_{\epsilon_y} \sum_{i=1}^n (\text{diag}(\bm X_i) \bm B_i)^{\prime}\text{diag}\bm (\bm X_i) \bm B_i  \\
    \ell_{\gamma} =  \sigma^{-2}_{\epsilon_y} \sum^{n}_{i=1}  (\text{diag}(\bm X_i) \bm B_i)^{\prime} (\bm y_i - \bm \theta_i)
    \end{gathered}
\end{equation}
\noindent
where $\text{diag}(\bm X_i)$ is a $m^y_i \times m^y_i$ matrix from vector $\bm X_i(\boldsymbol{\tau}^y_{i} - \Delta)$.
    \end{enumerate}

    \item For $\ell = 1, \dots, L$, sample spatial random effects:
      \begin{enumerate}
        \item     \begin{equation}
    \label{post-mu-sofr}
    \begin{gathered}
    [\bm \theta_{\ell} \mid \cdots] \stackrel{}{\sim} N(Q^{-1}_{\theta_{\ell}} \ell_{\theta_{\ell}},  Q^{-1}_{\theta_{\ell}}) \\
    Q_{\theta_{\ell}} =   \sigma^{-2}_{\theta_{\ell}} \bm Q  +  \sigma^{-2}_{\epsilon_y}\bm I_n\\
    \ell_{\theta_{\ell}} =  \sigma^{-2}_{\epsilon_y} \sum^{n}_{i=1}  \bm G_i^{\prime} (\bm y_i - \bm X_i(\boldsymbol{\tau}^y_{i} - \Delta))
    \end{gathered}
\end{equation}
\noindent
    \end{enumerate}

\item For $\ell = 1, \dots, L$, sample the loading curve parameters:
\begin{enumerate}
    \item Sample $\left[\lambda_{g_{\ell}} \mid \cdots\right] \sim \operatorname{Gamma}(1,  (\bm{\phi}_{\ell}^{\prime} \bm\Omega_{\phi} \bm{\phi}_{\ell})/2)$.
    \item Sample $\left[\bm{\phi}_{\ell} \mid \cdots\right] \sim N\left(\bm{Q}_{\phi_{\ell}}^{-1} \ell_{\phi_{\ell}}, \bm{Q}_{\phi_{\ell}}^{-1}\right)$ conditional on $\bm{C}_{\ell} \bm{\phi}_{\ell}=0$, where

\begin{equation}
    \label{post-phi-sofr}
    \begin{gathered}
    \bm{Q}_{\phi_{\ell}} = \sigma_{\epsilon_y}^{-2}\sum^{n}_{i=1} \theta_{{\ell},i}^2 \bm{V}_i^{\prime}\bm{V}_i + \lambda_{g_{\ell}} \bm{\Omega_{\psi}} \\
    \bm{\ell}_{\psi_{\ell}} =\sigma_{\epsilon_y}^{-2}\sum^{n}_{i=1}  \theta_{{\ell},i} {\bm V_i^{\prime}}  [ \bm{y}_i  - \bm X_i(\boldsymbol{\tau}^y_{i} - \Delta) - \bm{V}_i (\sum_{{\ell}^{*} \neq {\ell}} \bm{\phi}_{{\ell}^{*}} \theta_{{\ell}^{*} i} ) ] 
    \end{gathered}
\end{equation}

$\bm{C}_{\ell}=$ $\left(\bm{f}_1, \ldots, \bm{f}_{{\ell}-1}, \bm{f}_{{\ell}+1}, \ldots, \bm{f}_{\ell}\right)^{\prime} \bm{V}=\left(\bm \phi_1, \ldots, \bm{\phi}_{{\ell}-1}, \bm{\phi}_{{\ell}+1}, \ldots, \bm \phi_{\ell}\right)^{\prime}$. See \citet{kowal_dynamic_2021} for details on conditioning.
\end{enumerate}

\item Sample the time lag:
\begin{enumerate}
    \item Sample $\left[\Delta \mid \cdots\right] \sim \operatorname{Categorical}(\pi_{\Delta=0}, \dots, \pi_{\Delta=21})$ where

\begin{equation}
    \label{post-phi-sofr}
    \begin{gathered}
    \pi_{\Delta=s} =  \frac{z_{\Delta=s}}{\sum_{s^{*}=0}^{21} z_{\Delta=s^*}} \\
    z_{\Delta=s} =  \prod_{i=1}^n  \frac{-1}{\sqrt{2\pi\sigma_{\epsilon_y}}} \exp \left( \frac{1}{2\sigma^{2}_{\epsilon_y}} \Vert \bm y_i(\boldsymbol{\tau}^y_{i}) - \text{diag}(\bm \gamma(\boldsymbol{\tau}^y_{i}))\bm X_i(\boldsymbol{\tau}^y_{i} - s) - \bm \theta_i(\boldsymbol{\tau}^y_{i})\Vert^2\right) .
    \end{gathered}
\end{equation}
If given log-densities $\{\log (z_{\Delta=s})\} $ instead of $\{z_{\Delta=s}\}$, may use Gumbel-max trick \citep{jang_categorical_2017}.

\end{enumerate}

    \item For $k = 1, \dots, K$, sample factor parameters:
    \begin{enumerate}
        \item     \begin{equation}
    \label{post-mu-sofr}
    \begin{gathered}
    [\mu_k \mid \bm y, \bm x, \bm \mu_{-k} , \bm \alpha, \cdots] \stackrel{}{\sim} N(Q^{-1}_{\mu_k} \ell_{\mu_k},  Q^{-1}_{\mu_k}) \\
    Q_{\mu_k} =  \sigma_{\mu}^{-2} + \sum_{i=1}^{n} \sigma_{\epsilon_y}^{-2} ( \bm{f}^{\gamma}_{k,i})^{\prime} \bm{f}^{\gamma}_{k,i} +  \sigma_{\epsilon_x}^{-2} \bm{f}_{k,i}^{\prime}\bm{f}_{k,i} \\
    \ell_{\mu_k} =  \sum^{n}_{i=1} \sigma_{\epsilon_y}^{-2}( \bm{f}^{\gamma}_{k,i})^{\prime} (\bm y_i - \bm \theta_i - \bm{f}^{\gamma}_{k,i}\alpha_{ki}  - \sum_{k^{*} \neq k}  \bm{f}^{\gamma}_{k,i} \beta_{ki}) + \\
    \sigma_{\epsilon_x}^{-2}\bm{f}_{k,i}^{\prime}(\bm{x}_i - \bm{f}_{k,i}\alpha_{ki} + \sum_{k^{*} \neq k} \bm{f}_{k^{*},i}\beta_{k^{*}i})
    \end{gathered}
\end{equation}
\noindent
where $\bm{f}^{\gamma}_{k,i} = (f_{k}(\tau^y_{i,1}) \times \gamma(\tau^y_{i,1}), \dots, f_{k}(\tau^y_{i,m^y_i} )\times \gamma(\tau^y_{i,m^y_i}))^{\prime}$ and $\bm{f}_{k,i} = (f_{k}(\tau^x_{i,1}), \dots, f_{k}(\tau^x_{i,m^x_i} ))^{\prime}$.
        \item    
\begin{equation}
    \label{post-theta-sofr}
    \begin{gathered}
    [\alpha_{ki} \mid \bm y, \bm x, \bm \mu , \bm \alpha_{-k,i}, \cdots] \stackrel{}{\sim} N(Q^{-1}_{\alpha_{ki}} \ell_{\alpha_{ki}},  Q^{-1}_{\alpha_{ki}}) \\
    Q_{\alpha_{ki}} =  \sigma_{\alpha_i}^{-2} + \sigma_{\epsilon_y}^{-2} ( \bm{f}^{\gamma}_{k,i})^{\prime} \bm{f}^{\gamma}_{k,i} +  \sigma_{\epsilon_x}^{-2} \bm{f}_{k,i}^{\prime}\bm{f}_{k,i} \\
    \ell_{\alpha_{ki}} =   \sigma_{\epsilon_y}^{-2}( \bm{f}^{\gamma}_{k,i})^{\prime} (y_i - \bm \theta_i - \bm{f}^{\gamma}_{k,i}\alpha_{ki}  - \sum_{k^{*} \neq k} \bm{f}^{\gamma}_{k,i} \beta_{ki}) + \\
    \sigma_{\epsilon_x}^{-2}\bm{f}_{k,i}^{\prime}(\bm{x}_i - \bm{f}_{k,i}\alpha_{ki} + \sum_{k^{*} \neq k} \bm{f}_{k^{*},i}\beta_{k^{*}i})
    \end{gathered}
\end{equation}

    \end{enumerate}

\item For $k = 1, \dots, K$, sample the loading curve parameters:
\begin{enumerate}
    \item Sample $\left[\lambda_{f_k} \mid \cdots\right] \sim \operatorname{Gamma}(1,  (\bm{\psi}_k^{\prime} \bm\Omega_{\psi} \bm{\psi}_k)/2)$.
    \item Sample $\left[\bm{\psi}_k \mid \cdots\right] \sim N\left(\bm{Q}_{\psi_k}^{-1} \ell_{\psi_k}, \bm{Q}_{\psi_k}^{-1}\right)$ conditional on $\bm{C}_k \bm{\psi}_k=0$, where

\begin{equation}
    \label{post-phi-sofr}
    \begin{gathered}
    \bm{Q}_{\psi_k} = \sigma_{\epsilon_y}^{-2}\sum^{n}_{i=1} {\bm W_i^{\gamma\prime}}{\bm W_i^{\gamma}} \beta_{k,i}^2  + \sigma_{\epsilon_x}^{-2}\sum^{n}_{i=1} \beta_{k,i}^2 \bm{W}_i^{\prime}\bm{W}_i + \lambda_{f_k} \bm{\Omega_{\psi}} \\
    \bm{\ell}_{\psi_k} = \sigma_{\epsilon_y}^{-2}{\bm W_i^{\gamma\prime}}\sum^{n}_{i=1} \beta_{k,i} (\bm y_i - \bm \theta_i - {\bm W_i^{\gamma}}\sum_{k^{*} \neq k} \beta_{k,i} \bm{\psi}_{k^*}) + \\\sigma_{\epsilon_x}^{-2}\sum^{n}_{i=1}  \beta_{k,i} {\bm W_i^{\prime}}  [ \bm{x}_i  - \bm{W}_i (\sum_{k^{*} \neq k} \bm{\psi}_{k^{*}} \beta_{k^{*} i} ) ] 
    \end{gathered}
\end{equation}
where ${\bm W_i^{\gamma}} = \text{diag}(\bm \gamma(\boldsymbol{\tau}^y_{i}))\bm W_i$.

$\bm{C}_k=$ $\left(\bm{f}_1, \ldots, \bm{f}_{k-1}, \bm{f}_{k+1}, \ldots, \bm{f}_K\right)^{\prime} \bm{w}=\left(\bm \psi_1, \ldots, \bm{\psi}_{k-1}, \bm{\psi}_{k+1}, \ldots, \bm \psi_K\right)^{\prime}$. See \citet{kowal_dynamic_2021} for details on conditioning.
\end{enumerate}

\item 
 Sample the Multiplicative Gamma Process parameters: given $\mu_k$ and $\alpha_{k, i}$
 \begin{enumerate}
     \item Sample $\left[\delta_{\mu_1} \mid \cdots\right] \sim \operatorname{Gamma}\left(a_{\mu_1}+\frac{K}{2}, 1+\frac{1}{2} \sum_{k=1}^K \eta_{\mu_k}^{(1)} \mu_k^2\right)$ and $\left[\delta_{\mu_{\ell}} \mid \cdots\right] \sim \operatorname{Gamma}\left(a_{\mu_2}+\right.$ $\left.\frac{K-\ell+1}{2}, 1+\frac{1}{2} \sum_{k=\ell}^K \eta_{\mu_k}^{(\ell)} \mu_k^2\right)$ for $\ell>1$ where $\eta_{\mu_{\ell}}^{(k)}=\prod_{h=1, h \neq k}^{\ell} \delta_{\mu_h}$.
     \item Set $\sigma_{\mu_k}=\prod_{\ell \leq k} \delta_{\mu_{\ell}}^{-1 / 2}$.
     \item Sample $\left[\delta_{\alpha_1} \mid \cdots\right] \sim \operatorname{Gamma}\left(a_{\alpha_1}+\frac{K(n-1)}{2}, 1+\frac{1}{2} \sum_{k=1}^K \eta_{\alpha_k}^{(1)} \sum_{i=2}^n \alpha_{k, i}^2 \zeta_{\alpha_{k, i}}\right)$ and $\left[\delta_{\alpha_{\ell}} \mid \cdots\right] \sim \operatorname{Gamma}\left(a_{\alpha_2}+\frac{(K-\ell+1)(n-1)}{2}, 1+\frac{1}{2} \sum_{k=\ell}^K \eta_{\alpha_k}^{(\ell)} \sum_{i=2}^n \alpha_{k, i}^2 \zeta_{\alpha_{k, i}}\right)$ for $\ell>1$ where $\eta_{\alpha_{\ell}}^{(k)}=\prod_{h=1, h \neq k}^{\ell} \delta_{\alpha_h}$.
     \item Set $\sigma_{\alpha_k}=\prod_{\ell \leq k} \delta_{\alpha_{\ell}}^{-1 / 2}$.
     \item  Sample $\left[\zeta_{\alpha_{k, i}} \mid \cdots\right] \stackrel{\text { indep }}{\sim} \operatorname{Gamma}\left(\frac{\nu_\alpha}{2}+\frac{1}{2}, \frac{\nu_\alpha}{2}+\frac{\alpha_{k, i}^2}{2 \sigma_{\alpha_k}^2}\right)$.
     \item Sample $\left[\zeta_{\alpha_{k, i}} \mid \cdots\right] \stackrel{\text { indep }}{\sim} \operatorname{Gamma}\left(\frac{\nu_\alpha}{2}+\frac{1}{2}, \frac{\nu_\alpha}{2}+\frac{\alpha_{k, i}^2}{2 \sigma_{\alpha_k}^2}\right)$
     \item Set $\sigma_{\alpha_{k, i}}=\sigma_{\alpha_k} / \sqrt{\zeta_{\alpha_{k, i}}}$.
 \end{enumerate}

 \item Sample remaining variance parameters $\sigma^{2}_{\epsilon_x}$, $\sigma^{2}_{\epsilon_y}$ , $\sigma^{2}_{\theta_{p}}$:

 \begin{enumerate}
     \item 
 Sample $[\sigma^{-2}_{\epsilon_x} \mid \cdots] \sim \operatorname{Gamma}\left(a_{\epsilon_x}+\frac{1}{2}\sum_{i=1}^n m^x_i, b_{\epsilon_x}+\frac{1}{2}\sum_{i=1}^n \Vert \bm x_i - \bm{F}_i\bm{\beta}_{i} \Vert^2 \right)$.
 \item 
 Sample $[\sigma^{-2}_{\epsilon_y} \mid \cdots] \sim \operatorname{Gamma}\left(a_{\epsilon_y}+\sum_{i=1}^n m^y_i, 
 b_{\epsilon_y}+\frac{1}{2}\sum_{i=1}^n \Vert \bm y_i(\boldsymbol{\tau}^y_{i}) - \tilde{\bm y_i}(\boldsymbol{\tau}^y_{i}) \Vert^2 \right)$,
 \noindent
 where $\tilde{\bm y_i}(\boldsymbol{\tau}^y_{i}) = \text{diag}(\bm \gamma(\boldsymbol{\tau}^y_{i}))\bm X_i(\boldsymbol{\tau}^y_{i} - \Delta) - \bm \theta_i(\boldsymbol{\tau}^y_{i})$.
 \item 
 Sample $[\sigma^{-2}_{\theta_{p}} \mid \cdots] \sim \operatorname{Gamma}\left(.01+\frac{n}{2}, .01+\frac{1}{2}\bm \theta_p^{\prime} \bm Q \bm \theta_p \right)$.
 \end{enumerate}

 \item 
Sample $a_{\mu_1}, a_{\mu_2}, a_{\alpha_1}, a_{\alpha_2}, \text{and } \nu_{\alpha}$ independently using slice sampler \citep{neal_slice_2003}. 
\end{enumerate}

\subsection{Additional Application Results}

\begin{figure}[]
\centering
\includegraphics[width=1\textwidth]{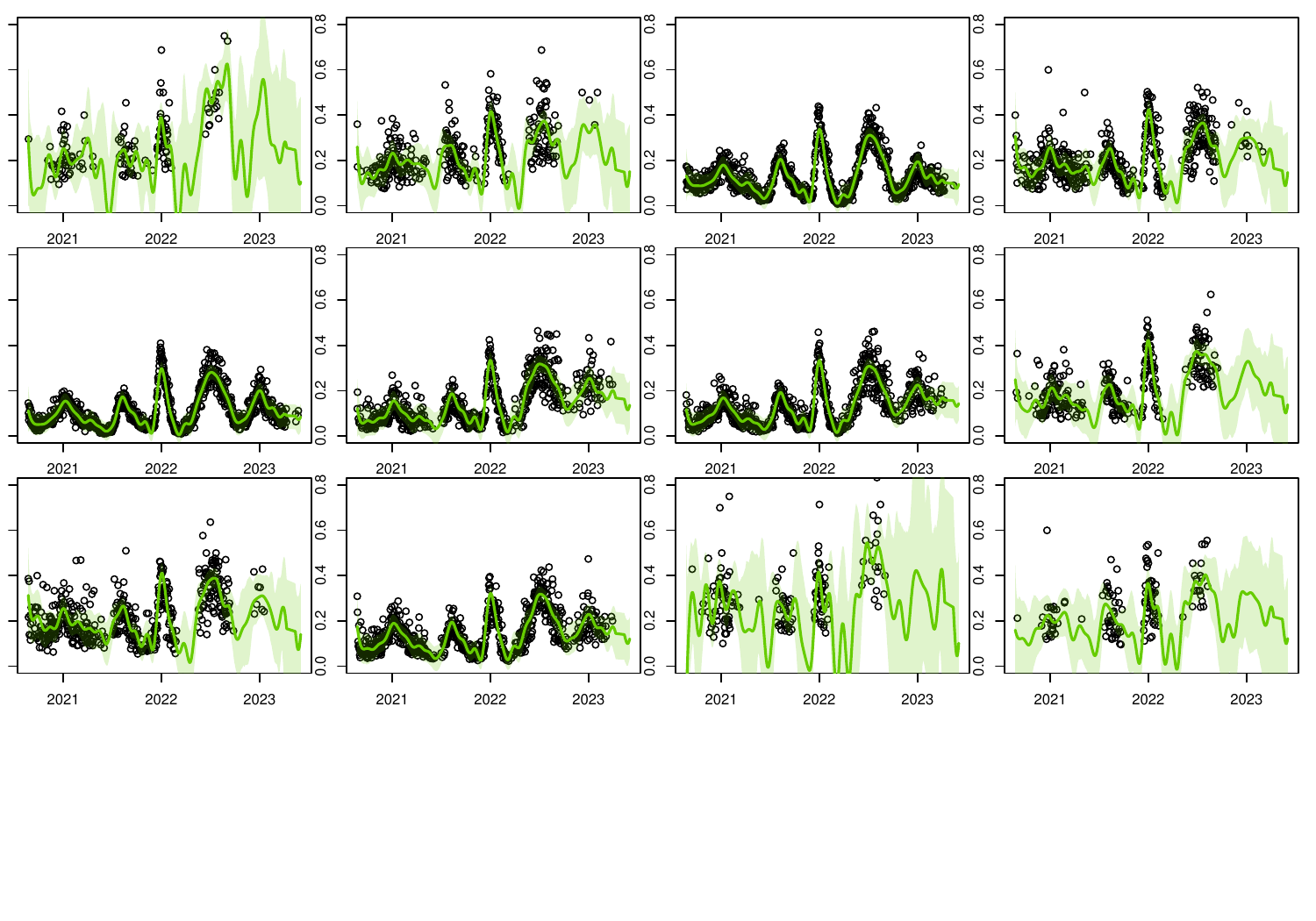}
\caption{Raw daily reported COVID-19 positivity rate values (dots) with fitted posterior mean curves (solid lines) and 95\% pointwise credible intervals (dashed lines) of the remaining 12 WWTPs. The functional regression model can provide predictions and uncertainty to missing portions of PR curves at locations with insufficient data.}
  \label{fig:estprpost2}
\end{figure}

\begin{figure}[]
\centering
\includegraphics[width=1\textwidth]{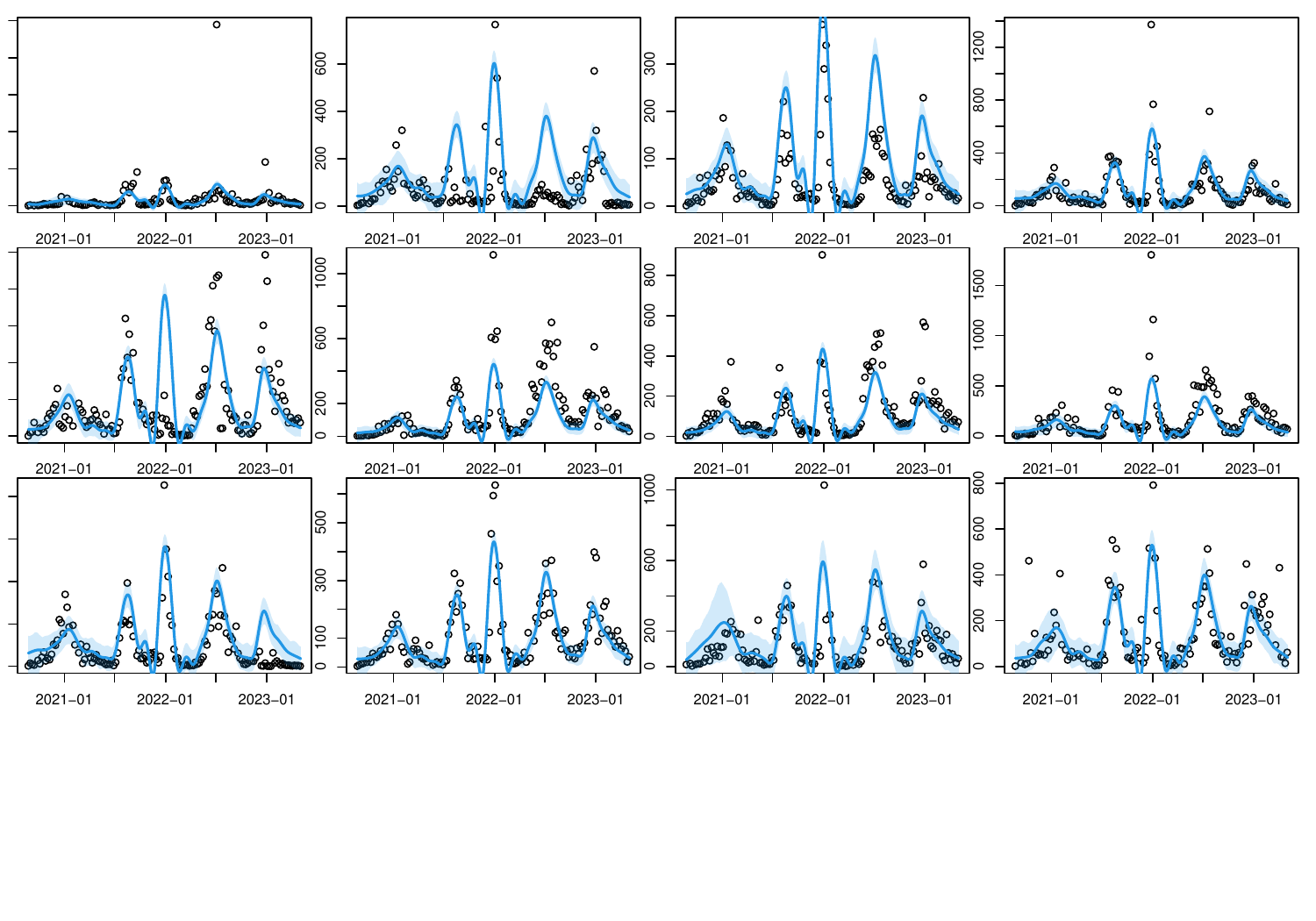}
\caption{Raw weekly SARS-CoV-2 wastewater measurements (dots) with fitted posterior mean curves (solid lines) and 95\% pointwise credible intervals (dashed lines) of the remaining 12 WWTPs. The Bayesian measurement error model provides estimates and uncertainty at times in between weekly measurements and incorporates information jointly from other model parameters.}
  \label{fig:estwwpost2}
\end{figure}

Figures \ref{fig:estprpost2}--\ref{fig:estwwpost2} show the raw values and fitted curves for the COVID-19 positivity rates and the SARS-CoV-2 wastewater measurements, respectively, for the remaining 12 WWTPs not shown in the main text.


\newpage

\bibliography{mybibww032624}